\documentclass[12pt,aps,prd,preprint,tightenlines,
superscriptaddress,
showpacs,nofootinbib]{revtex4-1}
\usepackage{amsmath,amssymb,amsthm,amsfonts}
\usepackage{slashed} 
\usepackage{graphicx}
\usepackage{subfigure}
\usepackage{dcolumn}
\usepackage{bm}
\usepackage{epsfig}
\usepackage{epstopdf}
\usepackage{setspace}
\usepackage{color}
\newcommand{\ben}{\begin{enumerate}}
\newcommand{\een}{\end{enumerate}}

\newcommand{\figref}[1]{Fig.~\ref{fig:#1}}

\newcommand{ \mysmall}[1]{\scriptscriptstyle #1} 

\newcommand{\be}{\begin{equation}}
\newcommand{\ee}{\end{equation}}
\newcommand{\bea}{\begin{eqnarray}}
\newcommand{\eea}{\end{eqnarray}}

\newcommand{\beq}{\begin{equation}}
\newcommand{\eeq}{\end{equation}}
\newcommand{\beqa}{\begin{eqnarray}}
\newcommand{\eeqa}{\end{eqnarray}}
\newcommand{\nn}{\nonumber}

\newcommand{\tanbeta}{\tan\beta}

\begin{document}
\title{Charged Slepton Flavor post the 8~TeV LHC:\\ 
A Simplified Model Analysis of Low-Energy Constraints\\ and LHC SUSY Searches}
\author{Lorenzo Calibbi}
\affiliation{Service de Physique Th\'eorique, Universit\'e Libre de Bruxelles,
1050 Brussels, Belgium}
\author{Iftah Galon}
\affiliation{Department of Physics and Astronomy, University of
California, Irvine, CA 92697, USA}
\author{Antonio Masiero}
\affiliation{Dipartimento di Fisica ed Astronomia ``G.Galilei'', Universit\`a degli Studi
di Padova, and Istituto Nazionale di Fisica Nucleare, Sezione di Padova,
Via Marzolo 8, 35131 Padova, Italy}
\author{Paride Paradisi}
\affiliation{Dipartimento di Fisica ed Astronomia ``G.Galilei'', Universit\`a degli Studi
di Padova, and Istituto Nazionale di Fisica Nucleare, Sezione di Padova,
Via Marzolo 8, 35131 Padova, Italy}
\author{Yael Shadmi}
\affiliation{Physics Department, Technion-Israel Institute of Technology, Haifa 32000, Israel}

\preprint{ULB-TH/15-02}

\begin{abstract}
Motivated by the null results of LHC searches, which together with the
Higgs mass, severely constrain minimal supersymmetric extensions
of the standard model, 
we adopt a model-independent approach to study charged slepton flavor. 
We examine a number of simplified models, with different subsets of 
sleptons, electroweak gauginos, and Higgsinos, and derive the allowed
slepton flavor dependence in the region probed by current LHC searches,
and in the region relevant for the 14~TeV LHC. 
We then study the impact of the allowed flavor dependence on
lepton plus missing energy searches. 
In some cases, flavor dependence 
significantly modifies the reach of the searches.
These effects may be even larger at the next LHC run,
since for the higher masses probed at 14~TeV,
larger flavor mixings and relative mass splittings are compatible
  with low-energy constraints. 
Retaining the full lepton flavor information can increase the sensitivity
of the searches.
\end{abstract}

\vskip1cm
\maketitle

\section{Introduction}
With the conclusion of the 8~TeV LHC run, supersymmetric extensions
of the Standard Model are greatly constrained by a variety of direct searches.
In the Minimal Supersymmetric Standard Model (MSSM),
squark masses are further constrained
by the 125~GeV Higgs mass, which requires either a large stop mixing or heavy stops.
In concrete models, the latter typically translate  
into lower bounds on the remaining squark masses. 
Thus, it is quite clear that the simplest scenarios, with all superpartners
near the TeV scale or below, are ruled out.
In particular, the direct production of sleptons, electroweak gauginos 
and Higgsinos may be the dominant signature of supersymmetry at the LHC.
More generally, it is conceivable that only some subset of superpartners
may be within reach, motivating a model-independent approach to
supersymmetry searches.

In this paper, we therefore adopt a simplified-model approach to
study charged slepton flavor.
There are several reasons why slepton flavor is interesting.
The origin of fermion masses is one of the most puzzling features of
the SM, hinting at some underlying flavor theory.
TeV-scale sleptons, if they exist, would provide a new portal
into the origin of flavor, both indirectly through Charged Lepton Flavor Violation
(CLFV), and through LHC measurements of their masses and couplings.
Even more importantly at this stage,
LHC slepton searches are in general sensitive to slepton flavor.
Thus for example, many slepton searches require Opposite Sign Same Flavor (OSSF) 
electron and muon pairs, assuming degenerate pure flavor states.
However, the slepton sector might feature a more generic flavor dependence, i.e.~non-degenerate 
masses of different flavors and/or mixing among flavor states.

The two main questions we will address are therefore:
\begin{enumerate}
\item What is the allowed slepton flavor dependence in the regions probed 
by current and future LHC searches?
\item How are these searches affected if such flavor dependence is indeed present?
\end{enumerate}

Apart from the fact that we want to examine the first question with as few
theory assumptions as possible, there are two other reasons for revisiting
it now. The first is very simple. As the LHC pushes the superpartner scale
to higher values, the allowed flavor mixings, and relative mass splittings
in the slepton spectrum can be larger, with potentially important effects
for LHC searches.
The second is again related to the measured Higgs mass.
As is well known, in the MSSM the strongest bounds on CLFV 
come from dipole transitions.
These are  enhanced in the presence of 
large Higgsino-gaugino mixing, and/or left-right slepton mixing,
since then the required chirality flip is supplied by the 
Yukawa vertex or by the slepton propagator.  
The measured Higgs mass therefore constitutes an important
input for CLFV. In some models, the 125~GeV Higgs mass favors
a large $\mu$ and heavy Higgsinos. 
If Higgsino diagrams
decouple because of a large $\mu$, 
the CLFV transitions have reduced contributions
and large slepton flavor dependence is possible.
In the following, we will therefore examine both scenarios
with active Higgsinos and scenarios with decoupled Higgsinos.
Finally, the use of simplified models will allow for a direct comparison with existing ATLAS and CMS analyses.

Indeed, current ATLAS and CMS electroweak searches~\cite{TheATLAScollaboration:2013hha,CMS:2013dea,Aad:2014nua,Aad:2014vma,Aad:2014iza,Khachatryan:2014qwa} already  
probe slepton masses up to a few hundred GeV in some cases.
Very roughly, searches based on two leptons (electrons and muons) and missing energy, 
extend to about 500~GeV chargino masses for a zero LSP mass, if charginos decay to substantially lighter
left-handed sleptons~\cite{Aad:2014vma}.
Searches based on three leptons are even more sensitive, as they can exclude neutralino/chargino masses up to
700 GeV for an  LSP mass below
about 350 GeV \cite{Khachatryan:2014qwa}.
Sleptons can also be directly produced via Drell-Yan processes with  $Z^0/\gamma^*$  s-channel exchange,
resulting in an opposite sign lepton-pair and missing energy. Because the relevant couplings are relatively 
small, these lead to the weakest bounds on the slepton and LSP masses. 
There is no bound for an LSP above $\approx$150$\div$200 GeV, and for a light LSP the bounds go up to 
a left-handed (right-handed) slepton mass of 300~GeV (250~GeV)~\cite{Aad:2014vma}. However, these bounds are very robust, 
as they only require the presence of a single slepton and the LSP. 

We will  consider several simplified models, including models used by
ATLAS and CMS
to interpret the searches for slepton electroweak production. 
Each of the models contains only a subset of the sleptons, gauginos and Higgsinos.
Schematically, the quantity constrained by CLFV bounds is the product
of the slepton relative mass splitting and the slepton mixing.
Since we are ultimately also interested in scenarios with large
mass splittings,  we calculate the CLFV observables in terms of the slepton
physical masses and mixings.
We then use these in Section~\ref{sec:SM_low} to 
derive the allowed regions in the slepton flavor parameters for each of 
the models in the limit of  small slepton mass splitting,
showing at the same time the limits set by direct LHC searches.

For each model, we also compute the predictions for the muon anomalous 
Magnetic Dipole Moment (MDM).
If the muon $g-2$ measurement~\cite{Bennett:2006fi,  Jegerlehner:2009ry,  HLMNT11,  DHMZ11} is interpreted as a deviation from the SM,
it requires, in the context of supersymmetry, light sleptons, gauginos and 
Higgsinos, with substantial $\tanbeta$ 
enhancement\footnote{As is well known, this discrepancy may be 
the result of hadronic SM 
contributions. For a recent review of experimental prospects for settling
this question see e.g.~\cite{Denig:2014mma}.}.
We note however that in the simple scenarios we discuss, 
the muon $g-2$ is related to the electron dipole 
moment by ``naive scaling'' with the fermion 
mass\footnote{See e.g.~\cite{Feng:2001sq,Giudice:2012ms} for a discussion of how this scaling can be violated
by flavor effects.}, 
and large values of $g-2$ require a solution of the supersymmetric CP problem.

We  then proceed to analyze the possible implications of lepton 
flavor violation for  LHC  
lepton plus missing energy searches,
in models with sleptons, Binos and Winos.
We consider DY slepton pair production, chargino pair production
and chargino-neutralino pair production. For each, we derive the excluded
region for models with non-degenerate sleptons, and for models with
some flavor mixing, and compare these to the flavor-blind results.

While we restrict ourselves to a model-independent approach, it
is important to stress that slepton flavor violation of the types we consider 
can arise in concrete and predictive 
models~\cite{Feng:2007ke,Kribs:2007ac,Nomura:2008pt,Nomura:2008gg,Shadmi:2011hs,Craig:2012yd,Calibbi:2012yj,Craig:2012di,Calibbi:2014yha}. 
Indeed, any mechanism which explains fermion masses is likely to control
also sfermion masses. This has been utilized in different frameworks
to obtain flavor-dependent spectra consistent with CLFV bounds.
In particular, scenarios with large mass splittings can be compatible with CLFV
constraints in alignment models~\cite{nirseiberg}, in which some mechanism, such as
flavor symmetries, suppresses flavor mixing~\cite{Feng:2007ke,Shadmi:2011hs,Abdullah:2012tq,
Galon:2013jba,Calibbi:2014pza}.

This paper is organized as follows.
In Section~\ref{sec:setup}, we set the notation for the slepton flavor parameters. 
In Section~\ref{sec:dipoles} we introduce the low-energy observables related to leptonic dipoles,
and review the current experimental sensitivities as well as future prospects.
In Section~\ref{sec:SM_low}, we analyze the low-energy flavor constraints for each of the
models. 
For reference, we show these constraints together with the
limits on flavor-blind sleptons from direct LHC searches. 
We then turn to the signatures of flavor-dependent  models at the LHC,
and reinterpret several analyses in terms of flavor dependent slepton spectra
in Section~\ref{sec:SM_high}.
We conclude with some remarks in Section~\ref{sec:conclusions}.
Finally, the supersymmetric expressions for the dipole amplitudes are collected in the Appendix.

 
\section{General setup: slepton flavor parameters}
\label{sec:setup}
We begin by explaining our conventions and assumptions.
In each of the models we consider, we assume a single dominant source of flavor violation,
so that the main signatures of interest can be described using two slepton states.
We use $L$ ($R$) to denote ``left-handed'' (``right-handed'') sleptons. 
We will mostly assume small LR mixing, 
so that the two sleptons are predominantly $L$ or $R$.

Working in the fermion mass basis, with diagonal gaugino-slepton-lepton couplings, we then write the slepton 
mass matrices  as
\begin{equation}
M^{2}_{LL} = 
\begin{pmatrix}      
m^2_{L_1} & \Delta^{12}_{LL} \\
\Delta^{21}_{LL} & m^2_{L_2} 
\end{pmatrix}~,\qquad\qquad
M^{2}_{RR} = 
\begin{pmatrix}      
m^2_{R_1} & \Delta^{12}_{RR} \\
\Delta^{21}_{RR} & m^2_{R_2} 
\end{pmatrix}~,
\end{equation}
which can be diagonalized through unitary matrices $U_L$ and $U_R$, respectively, defined as
\begin{equation}
U^{\dagger}_L M^{2}_{LL} U_L = {\rm diag}(m^2_{\widetilde \ell_1},m^2_{\widetilde \ell_2})~,\qquad
U^{\dagger}_R M^{2}_{RR} U_R = {\rm diag}(m^2_{\widetilde e_1},m^2_{\widetilde e_2})\,,
\label{eq:U_LR}
\end{equation}
where $U_L$ and $U_R$ read
%
%
%
\begin{equation}
U_L = \begin{pmatrix} \cos\theta_{\mysmall L} & -\sin\theta_{\mysmall L} \\
\sin\theta_{\mysmall L} & \cos\theta_{\mysmall L} \end{pmatrix}~, ~~~
U_R = \begin{pmatrix} \cos\theta_{\mysmall R} & -\sin\theta_{\mysmall R} \\
\sin\theta_{\mysmall R} & \cos\theta_{\mysmall R} \end{pmatrix}~,
\end{equation}
where, for simplicity, we have assumed CP conservation, i.e. $\Delta^{12}_{LL}=\Delta^{21}_{LL}$ and 
$\Delta^{12}_{RR} = \Delta^{21}_{RR}$. 
EDMs in these scenarios thus only arise from ``flavor-diagonal'' phases\footnote{The contribution to EDMs from flavor-changing
parameters has been discussed in \cite{Feng:2001sq,Hisano:2007cz,Hisano:2008hn}.}.
The flavor mixing angles $\sin\theta_{L,R}$, $\cos\theta_{L,R}$ 
are defined as
\be
\sin\theta_{\mysmall L}\cos\theta_{\mysmall L}=
\frac{\Delta^{21}_{LL}}{(m_{\widetilde{\ell}_1}^{2}-m_{\widetilde{\ell}_2}^{2})}\,,
\qquad
\sin\theta_{\mysmall R} \cos\theta_{\mysmall R}=
\frac{\Delta^{21}_{RR}}{(m_{\widetilde{e}_1}^{2}-m_{\widetilde{e}_2}^{2})}\,,
\label{eq:mixing_angles}
\ee
where $\widetilde\ell_1$ and $\widetilde e_1$ are the heaviest mass eigenstates.

We will often use the average slepton mass-squared, $m^2_M$, and the mass splitting $\Delta m_M$,
given by,
\bea
m^2_L &\equiv& (m^2_{\widetilde\ell_1}+ m^2_{\widetilde\ell_2})/2\,,\qquad \Delta m_L = m_{\widetilde\ell_1} - m_{\widetilde\ell_2}\,,
\\
m^2_R &\equiv& (m^2_{\widetilde e_1}+ m^2_{\widetilde e_2})/2\,,\qquad \Delta m_R = m_{\widetilde e_1} - m_{\widetilde e_2}\,.
\eea
It is then useful to define the dimensionless MIs as 
\beq
\delta^{21}_{LL} \equiv \frac{\Delta^{21}_{LL}} {m^2_{L}} 
\,,
\qquad\qquad
\delta^{21}_{RR} \equiv \frac{\Delta^{21}_{RR}} {m^2_{R}} 
\label{eq:deltas1}
\eeq
In the limit of small mass splitting,
\beq
\delta^{21}_{LL} 
\approx \frac{\Delta m_L}{m_L}\sin2\theta_L\,,
\qquad\qquad
\delta^{21}_{RR} 
\approx\frac{\Delta m_R}{m_R}\sin2\theta_R\,.
\label{eq:deltas2}
\eeq
As we will discuss in detail in the following sections, 
while LHC searches are sensitive to the 
slepton masses and mixings separately, CLFV processes essentially constrain the product 
of the mixing and relative mass splitting, i.e. $\delta_{LL}$ and $\delta_{RR}$.
These can be small either because the mass splittings are small, or because the mixing is small,
as in alignment models~\cite{nirseiberg}. As shown in~\cite{Raz:2002zx, Nir:2002ah} the MIA gives a good estimate
of CLFV constraints even in this latter case. However, for detailed studies of LHC processes
with large mass splittings and small mixings we will employ the full expressions for the dipole
amplitudes.

For simplicity, 
we suppress the $L,R$ indices on  $\Delta m$ and $\theta$, 
whenever only a single mass splitting and a single mixing angle are present.

In some of the models, we also consider left-right slepton mixing.
Generically, this mixing is given by a $3\times3$ matrix, of the form 
$y_\ell(A-\mu\tan\beta)$,
where $y_\ell$ is the lepton Yukawa matrix.
We will neglect the $A$-terms in the flavor-diagonal left-right mixings,
assuming that  the main contribution is due to the  $\mu\tan\beta$ term,  
so that the mixing is  proportional to the relevant lepton mass.
We will consider, however, $A$-term-induced flavor violation encoded in
\beq
\delta^{21}_{LR} \equiv \frac{m_{\ell_2} A_{21}} {\sqrt{m^2_{L}m^2_{R}}} 
\,,
\qquad\qquad
\delta^{21}_{RL} \equiv \ \frac{m_{\ell_1} A_{12}} {\sqrt{m^2_{L}m^2_{R}}}\,. 
\label{eq:deltasLR}
\eeq

\section{Leptonic dipoles and low energy observables}\label{sec:dipoles}
The search for flavor violation in charged leptons is certainly one of the most interesting goal of flavor physics 
in the  near future. 
Indeed, neutrino oscillations have shown that lepton flavor is not conserved, and TeV-scale New Physics ({\small NP}) can lead 
to observable CLFV. Among the most interesting CLFV channels are $\mu\to e\gamma$, $\mu \to eee$, $\mu \to e$
conversion in Nuclei as well as $\tau$ LFV processes.
The current status and future experimental sensitivities for LFV processes as well as the electron EDM 
are collected in Table~\ref{nextgenexp}.
\begin{table}[t!]
\centering
\setlength{\extrarowheight}{2.5pt}
\begin{tabular}{|c|r|r|}
\hline
LFV Process & ~Present Bound~ & ~Future Sensitivity~  \\
\hline
$\mu^+ \to e^+ \gamma$ & $5.7 \times 10^{-13}$ \cite{Adam:2013mnn} & $\approx 6 \times 10^{-14}$ \cite{Baldini:2013ke}  \\
$\mu^+ \to e^+e^+e^-$ & $1.0 \times 10^{-12}$\cite{Bellgardt:1987du} & $\mathcal{O}(10^{-16})$ \cite{Blondel:2013ia}\\
$\mu^-$ Au $\to$ $e^-$ Au & $7.0 \times 10^{-13}$ \cite{Bertl:2006up} & $ ?~~~~~~~~~\quad $  \\
$\mu^-$ Ti $\to$ $e^-$ Ti & $4.3 \times 10^{-12}$ \cite{Dohmen:1993mp} & $?~~~~~~~~~\quad$ \\
$\mu^-$ Al $\to$ $e^-$ Al & $-~~~~~~\quad$  & $\mathcal{O}(10^{-16})$ \cite{comet,mu2e} \\
$\tau^\pm \to \mu^\pm \gamma$ & $4.4 \times 10^{-8}$ \cite{Aubert:2009ag}& $10^{-8}\div10^{-9}$ \cite{Hayasaka:2013dsa} \\
$\tau^\pm \to \mu^\pm\mu^+\mu^-$ & $2.1\times10^{-8}$\cite{Hayasaka:2010np} & $10^{-9}\div 10^{-10}$ \cite{Hayasaka:2013dsa}  \\
\hline
Electron EDM & ~Present Bound~ & ~Future Sensitivity~  \\
\hline
$d_e ({\rm e~cm})$ & $8.7 \times 10^{-29}$ \cite{Baron:2013eja} & $?~~~~~~~~~\quad$ \\
\hline
\end{tabular}
\caption{Current experimental bounds and future sensitivities for some low-energy LFV
         observables and the electron EDM.}
\label{nextgenexp}
\end{table}

In supersymmetric extensions of the SM,  new sources of CLFV stem from 
the soft SUSY-breaking sector
since the lepton and slepton mass matrices are generally misaligned \cite{Borzumati:1986qx,Hisano:1995cp}.
The dominant CLFV effects are captured by the dipole operators, 
\begin{equation}
\mathcal L = e\frac{m_{\ell}}{2}
{\bar\ell}_{i}\sigma_{\mu\nu} F^{\mu\nu}
\left(A_{L}^{ij} P_L + A_{R}^{ij} P_R\right)\ell_{j}
\qquad i,j = e,\mu,\tau\,,
\label{eq:eff_lagr_LFV}
\end{equation}
which arise from sneutrino-chargino and slepton-neutralino loops.
The Lagrangian~\eqref{eq:eff_lagr_LFV} leads to,
\begin{equation}
\frac{{\rm BR}(\ell_i \to \ell_{j}\gamma)}{{\rm BR}(\ell_i \to \ell_{j} \nu_{i} \bar\nu_{j})} =
\frac{48\pi^3\alpha}{G^2_F}
\left( 
|A_{L}^{ij}|^2 + |A_{R}^{ij}|^2
\right)\,,
\label{eq:BR_LFV}
\end{equation}
and the following model-independent relations  hold:
\begin{eqnarray}
\frac{{\rm BR}(\ell_i\rightarrow \ell_j\ell_k\bar{\ell}_k)}{{\rm BR}(\ell_i\rightarrow \ell_j\bar{\nu_j}\nu_{i})}
&\simeq&
\frac{\alpha_{el}}{3\pi}
\bigg(\log\frac{m^2_{\ell_i}}{m^2_{\ell_k}}-3\bigg)
\frac{{\rm BR}(\ell_i\rightarrow \ell_j\gamma)}{{\rm BR}(\ell_i\rightarrow \ell_j\bar{\nu_j}\nu_{i})}~,
\nonumber\\
{\rm CR}(\mu\to e~\mbox{in N})
&\simeq&
\alpha_{\rm em} \times {\rm BR}(\mu\rightarrow e\gamma)~.
\label{eq:dipole}
\end{eqnarray}
As a result, the current MEG bound ${\rm BR}(\mu\to e\gamma)\sim 5\times 10^{-13}$
already implies that ${\rm BR}(\mu\to eee)\leq 3\times 10^{-15}$ and
${\rm CR}(\mu\to e~\mbox{in N})\leq 3\times 10^{-15}$.

The CLFV transitions are tightly related to the magnetic and electric leptonic dipole moments, 
which are given by the effective Lagrangian of Eq.~(\ref{eq:eff_lagr_LFV}) with $\ell_i=\ell_j$. 
Denoting the anomalous magnetic moments by $\Delta a_{\ell}$, and the leptonic EDMs  by 
$d_\ell$, we can write them as
\begin{equation}
\Delta a_{\ell_i} = m^{2}_{\ell_i}~{\rm Re} \left( A_L^{ii} + A_R^{ii} \right)\,,\qquad
\frac{d_{\ell_i}}{e} = \frac{m_{\ell_i}}{2}~{\rm Im} \left( A_L^{ii} + A_R^{ii} \right)\,.
\end{equation}
Both $\Delta a_{\ell_i}$ and $d_{\ell_i}$ are extremely sensitive probes of new physics.
In particular, the current anomaly $a_\mu = (g-2)_\mu /2$ which exhibits
a $\sim 3.5 \sigma$ discrepancy between the {\small SM} prediction and the
experimental value~\cite{Jegerlehner:2009ry}
$\Delta a_\mu =a_\mu^{ \rm EXP}-a_\mu^{ \rm SM} = 2.90 \,(90)\times 10^{-9}$,
reinforces the expectation of detecting $\mu \to e\gamma$, hopefully within
the MEG resolutions.
In concrete {\small NP} scenarios, $\Delta a_{\ell}$, $d_{\ell}$ and
${\rm BR}(\ell\to \ell^{\prime}\gamma)$ are expected to be correlated.
However, their correlations crucially depend on the unknown flavor and
CP structure of the {\small NP} couplings.

We now review the main features of the superpartner contributions to the 
Lagrangian~(\ref{eq:eff_lagr_LFV}).
The chiral symmetry breaking source required by the dipole transition can be implemented 
in three different ways:
(i) through a chirality flip on the external fermion line, 
(ii) through mixing effects in the chargino/neutralino mass matrices, 
or (iii) through LR or RL mixings in the charged-slepton mass matrix. 
In~(i), the amplitudes are independent of
$\tanbeta$, while in~(ii) the leading effects are proportional to $\tanbeta$ 
because of the lepton Yukawa coupling at 
the Higgsino-lepton-slepton vertex. 
In~(iii), the amplitudes are proportional to the LR/RL mixing $\sim A -\mu\tan\beta$ 
and therefore grow with $\mu\tan\beta$.

In the next section we present the bounds on $\delta_{MN}$
in the limit of degenerate slepton masses 
(for earlier works, see \cite{Gabbiani:1988rb,Gabbiani:1996hi,Masina:2002mv,Paradisi:2005fk}).
For this purpose, a computation in the
so-called Mass Insertion Approximation (MIA) would be sufficient.
However, since we are interested also in scenarios with large mass splittings (and small
mixings), a full computation in the mass-eigenstate basis is unavoidable. 
In the Appendix we provide very compact expressions 
for the $\ell_i \to \ell_j \gamma$ amplitudes, distinguishing among the ways in which the chirality 
flip is implemented. We also collect the expressions for the muon $g-2$, and the electron EDM. 
In order to simplify the expressions as much as possible while keeping all the 
important features, these are obtained by treating SU(2) breaking effects in the 
chargino/neutralino mass-matrices 
as perturbations~\cite{Paradisi:2005fk}, 
and working within a two family framework. 

\begin{figure}[t]
\centering
\includegraphics[scale=0.6]{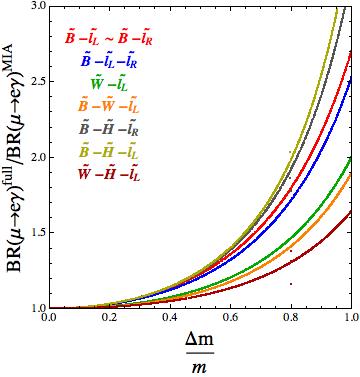}
\caption{
The full vs.~MIA results for ${\rm BR}(\mu\to e\gamma)$ in the simplified 
models considered 
in this paper as a function of the normalized mass-splitting $\Delta m/m$.}
\label{fig:MI_vs_full}
\end{figure}

For completeness, we also show in the Appendix the MIA amplitudes for $\ell_i \to  \ell_j \gamma$.
In order to appreciate the limit of validity of the MIA results compared to the 
full results in the mass-eigenstates,  
we plot in Fig.~\ref{fig:MI_vs_full} the ratio 
${\rm BR}(\mu\to e\gamma)^{\rm full} / {\rm BR}(\mu\to e\gamma)^{\rm MIA}$
for the different simplified models we will discuss in the following, 
as a function of the normalized mass-splitting $\Delta  m/m$ 
where $m$ is the average slepton mass. 
As we can see, the two calculations are completely equivalent in the limit $\Delta m/m\to0$. Moreover, 
the MIA results are still reasonably accurate up to mass splitting of order 
$\Delta m/m \lesssim 0.5$,
while they underestimate the result for larger mass splittings.

\section{Simplified models: LFV versus LHC bounds}
\label{sec:SM_low}
In this Section, we analyze the implications of the current CLFV bounds for 
different simplified models, and display the excluded regions together
 with the results of LHC searches for sleptons and charginos/neutralinos.
The latter assume flavor-blind sleptons, with degenerate selectrons and smuons,
and no flavor mixing\footnote{A notable exception is \cite{atlasaux}, where separate
limits on the selectron and smuon masses are shown in the auxiliary plots.}.
In this section, we simply display the limits from CLFV experiments together with the LHC limits.
In the next section, we discuss the possible effects of relaxing the assumption of flavor blind sleptons,
and address the impact of large inter-generation mixing, 
or mass splittings, on LHC searches.

For simplicity, we restrict ourselves to models defined by at most three mass scales.
We denote each model by the light superpartners it contains. For example, 
in $\widetilde{\ell}_R \widetilde B$ models, the only superpartners are 
right-handed sleptons and a Bino-like lightest neutralino.
All other sleptons, neutralinos and charginos are assumed to be very heavy, so that they are 
beyond the reach of the LHC, and furthermore, their contributions to 
the various dipole transitions 
can be neglected. The latter is a much stronger assumption. 
Indeed, the cross sections for producing 
heavy superpartner pairs fall very fast with the superpartner mass, whereas the contributions 
of heavy superpartners to CLFV processes decouple more slowly. We will address this point in 
detail at the end of this Section, and show the parameter ranges for which the simplified 
expressions of each model represent 
a good approximation of the full amplitude of the CLFV processes.

We focus here on LHC searches for leptons plus missing energy,
which require a neutralino LSP~\cite{TheATLAScollaboration:2013hha,CMS:2013dea,Aad:2014nua,Aad:2014vma,Aad:2014iza,Khachatryan:2014qwa}. 
Different hierarchies are possible of course, with the charged slepton NLSP 
decaying to a gravitino
or through R-parity violating couplings.
The LHC signatures then depend on the NLSP lifetime and decay products.
Thus for example, a single long-lived, left-handed slepton is excluded for masses below
339~GeV based only on its Drell-Yan production~\cite{Chatrchyan:2013oca}.
From this, the direct production bound on two (three) degenerate slepton 
flavors can be estimated
to be 400 (435)~GeV ~\cite{Calibbi:2014pza}.
Flavor effects in such scenarios were studied for example in~\cite{Feng:2007ke,
Feng:2009yq,Feng:2009bd}.

\subsection{$\widetilde{\ell}_L\widetilde B$ models}
We begin with one of the simplest models, with only the left-handed 
sleptons and a Bino neutralino.  
This model is a good starting point for understanding some of the main features of the 
flavor-collider interplay. 
On the one hand, the left-handed sleptons have larger Drell-Yan production cross-sections 
compared to the right-handed sleptons. Consequently, LHC 
searches have a higher reach for left-handed slepton masses. On the other hand, 
the couplings of left-handed sleptons to the Bino 
are a factor of 2 smaller than the couplings of right-handed sleptons. 
The left-handed slepton masses are therefore less constrained 
by flavor measurements. 

The various dipole amplitudes are very simple in this case. 
Using the expressions collected in the appendices,
\bea
A_L &=&  (A^{n_1}_{L})_{\mysmall U(1)}\,,\qquad A_R \simeq 0\,.
\nonumber\\
\Delta a_\mu &=& \left(\Delta a^{n_1}_{\mu}\right)^{\!\mysmall L}_{\!\mysmall U(1)}\,,
\qquad
d_e \simeq 0
\,.
\label{eq:lLB}
\eea
Thus for example, for small slepton mass splitting, the amplitude for $\mu\to e\gamma$ reads
\bea
A_L=\frac{\alpha_Y}{4\pi} \frac{\delta_{LL}^{21}}{m_L^2} f_{1n} (x_{\mysmall 1L})\,.
\label{eq:lLB2}
\eea

A few features of this model are worth stressing. 
First, the electron EDM, $d_e$, vanishes. 
The required chirality flip can only occur on an external fermion line. 
The two 
Bino-lepton-slepton loop-vertices are therefore complex-conjugates of each other. 
Consequently, the loop amplitude is real and the EDM vanishes. 
Second, as $\Delta a_\mu$ is always negative, 
the muon $g-2$ anomaly cannot be accounted for in this scenario. 
Furthermore, the contribution is numerically small, so that the model predicts 
a SM-like muon $g-2$.

Third,  and most importantly for our purposes, it is straightforward to compare 
the reach of direct LHC lepton plus
 missing energy searches~\cite{TheATLAScollaboration:2013hha,CMS:2013dea,Aad:2014nua,Aad:2014vma,Aad:2014iza,Khachatryan:2014qwa} 
to  $\ell_i\to \ell_j \gamma$ constraints in this case.
Since we assume that the Higgsinos are decoupled, and that the only light sleptons are
purely left-handed, the relevant LHC signatures as well as the LFV constraints are  determined
solely by the Bino mass $M_1$,
and, in the limit of flavor blind slepton masses, the slepton mass $m_L$.

\begin{figure}[t]
\centering
\includegraphics[scale=0.6]{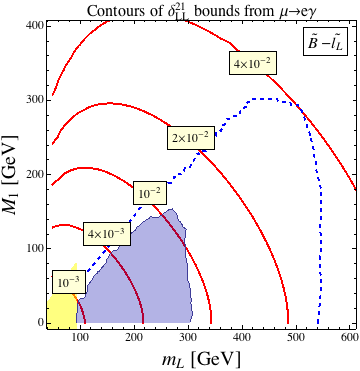}
\hspace{0.5cm}
\includegraphics[scale=0.6]{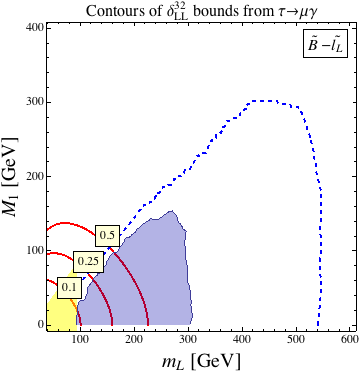}
\caption{
Upper limits on $\delta^{21}_{LL}$ (left) and $\delta^{32}_{LL}$ (right) in the plane of the Bino
mass, $M_1$, and common L-slepton mass, $m_L$ for 
the model $\widetilde \ell_LB$. 
The light-blue area is excluded by the ATLAS~\cite{TheATLAScollaboration:2013hha} direct search (assuming flavor blind
sleptons), 
the yellow region refers to the LEP exclusion. The dashed line refers
to the future LHC limit with $\sqrt{s}$=14 TeV and  
$\mathcal{L}=100~\rm fb^{-1}$, as estimated in
\cite{Eckel:2014dza}.}
\label{fig:model_A}
\end{figure}

In Fig.~\ref{fig:model_A}, we show the region excluded by the ATLAS search~\cite{TheATLAScollaboration:2013hha}\footnote{We choose to show the results of the ATLAS preliminary analysis~\cite{TheATLAScollaboration:2013hha} instead of those of the published paper~\cite{Aad:2014vma} for the sake of consistency with the numerical
results, cf.~the next Section. We notice however that in terms of limits on the slepton-neutralino mass plane, the two analyses are practically equivalent.}, 
which assumes
flavor blind sleptons, together with
the constraints from $\mu\to e\gamma$ (left panel) or $\tau\to \mu\gamma$ (right panel)
in the $(m_L,~M_1)$ plane. As noted above, in the absence of flavor dependence, 
these are the physical masses of the Bino and sleptons in this model.
%
%
%
%
The contours correspond to the upper bounds on $\delta^{21}_{LL} \equiv \Delta_{LL}^{21}/m^2_L$ 
(left) and $\delta^{32}_{LL} \equiv \Delta_{LL}^{32}/m^2_L$ (right),
obtained using the latest limits on ${\rm BR}(\ell_i\to \ell_j \gamma)$ 
listed in Table~\ref{nextgenexp}.
The yellow region represents the LEP exclusion. 
The light-blue area is excluded by the ATLAS search for Drell-Yan slepton pair production, 
with each slepton decaying to a Bino plus lepton,
leading to 
two Opposite Sign, Same Flavor (OSSF) leptons ($e^+e^-$ and $\mu^+\mu^-$) 
plus missing transverse 
momentum~\cite{TheATLAScollaboration:2013hha}.
Note that this is the only possible channel for slepton production in these models
for $M_L>M_1$.
For $m_L$ above or near $M_1$, the LHC signatures of the model are qualitatively different, 
and depend in particular on the identity and mass of the LSP, which determine the slepton
lifetime.
Thus for example, three mass-degenerate long-lived left-handed sleptons are excluded up to 430~GeV~\cite{Chatrchyan:2013oca,Calibbi:2014pza}. 

We see that in the light-blue region probed by the LHC, the allowed flavor dependence
can be substantial. There are essentially no constraints on the stau-smuon system, and
even in the selectron-smuon system,  $\delta^{21}_{LL}$ at the percent level is allowed.
We also show here (blue dashed line) the projected 
95\% CL exclusion limit at the $\sqrt{s}$=14 TeV LHC for $\mathcal{L}=100~\rm fb^{-1}$, 
as estimated in \cite{Eckel:2014dza}. 
Naturally, the allowed flavor dependence is this higher mass range is even larger.

%
\subsection{$\widetilde{\ell}_L\widetilde W$ models}
%
\begin{figure}[t!]
\centering
\includegraphics[scale=0.6]{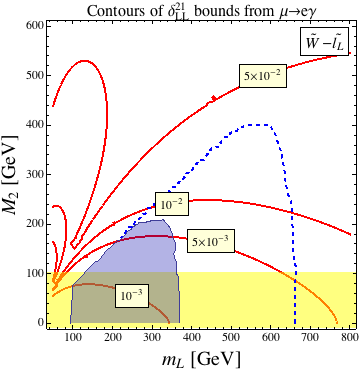}
\hspace{0.5cm}
\includegraphics[scale=0.6]{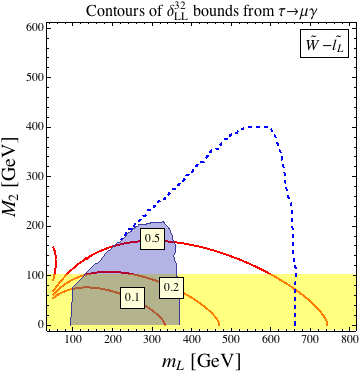}
\caption{
Upper limits on $\delta^{21}_{LL}$ (left) and $\delta^{32}_{LL}$ (right) in the plane $(m_L,~M_2)$ 
for the model~$\widetilde{\ell}_L\widetilde W$. The yellow region refers to the LEP exclusion.
The light-blue area (dashed line) represents the current (future) LHC  exclusion, as estimated in \cite{Eckel:2014dza}.}
\label{fig:model_D}
\end{figure}
%
We now assume that only the Winos and left-handed sleptons are light.
The light spectrum is given by nearly degenerate, Wino-like chargino and neutralino,
as well as 
charged sleptons and sneutrinos. 
Unlike the previous model, here the sneutrinos
play a role in both the dipole amplitudes and in the LHC processes of interest,
and we will assume that the masses of the charged sleptons and sneutrinos
are very close, with mass-squared differences less than $M_W^2$, as is the case in the MSSM.

The dipole amplitudes are again quite simple, with the chirality flip occurring on
the external fermion line(s). The expressions can be obtained 
from~Eqs.~(\ref{eq:lLB}, \ref{eq:lLB2}) 
with $\alpha_Y\to \alpha_2$,
\beqa
\label{eq:2b}
A_{L} &=& (A^{n_1}_{L})_{\mysmall SU(2)} + (A^{c_1}_{L})_{\mysmall SU(2)}\,,
\quad\qquad\qquad
A_{R} \simeq 0\,,
\\
\Delta a_\mu &=& \left(\Delta a^{n_1}_{\mu}\right)_{\!\mysmall SU(2)} + \left(\Delta a^{c_1}_{\mu}\right)_{\!\mysmall SU(2)}\,,
\qquad\qquad
d_e \simeq 0
\label{eq:amu2b}
\,. 
\eeqa

Again, 
the LHC signatures of these models are largely determined by the identity of the LSP, and by the
mass splitting between the LSP and NLSP. Thus for example, for an LSP neutralino and an 
almost degenerate NLSP chargino, chargino masses up to $\sim300-500$~GeV are excluded as 
the chargino-neutralino mass difference varies between $\sim 160-140$~MeV by searches for disappearing tracks~\cite{Aad:2013yna}.

Slepton pair production followed by decays to leptons plus Winos was recently
studied in these models in~\cite{Eckel:2014dza}, by recasting the ATLAS analysis~\cite{Aad:2014vma}. 
In Fig.~\ref{fig:model_D}, we show the estimates of~\cite{Eckel:2014dza} for the region excluded
by current LHC data (light-blue area),  and for the reach of the 14~TeV LHC (dashed line),
assuming a flavor-blind slepton spectrum, 
for different choices of the Wino mass $M_2$ and the common slepton mass $m_L$.
We also plot the upper bounds on $\delta^{21}_{LL}$ (left) and 
$\delta^{32}_{LL}$ (right), 
and the LEP limit, $m_{\widetilde{\chi}^\pm_1} > 103$ GeV 
(assuming $m_{\widetilde{\chi}^0_1}= m_{\widetilde{\chi}^\pm_1}= M_2$). 
The present LHC exclusion, as estimated in~\cite{Eckel:2014dza}, is stronger 
than in the Bino-LSP case, because  
of the larger number of production modes (such as sneutrino-slepton and sneutrino-sneutrino), 
that can lead to dilepton events. 
As for the LFV processes,
as we can see from Fig.~\ref{fig:model_D}, a novel 
feature of this model 
is the possibility of cancellations between the two 
contributions to $A_{L}$, cf.~Eq.~(\ref{eq:2b}). This occurs for $m_L \approx 1.5\times M_2$, 
for which the
LSP is a sneutrino. 
However, for the region of parameter space probed by current LHC lepton plus missing 
energy searches, the allowed flavor dependence is more constrained than in the 
$\widetilde{\ell}_L\widetilde B$ model.

Finally, $(g-2)_\mu$ is non-zero in this model,  
because of the  chargino contribution (see Eq.~(\ref{eq:amu2b})).
However, the resulting $\Delta a_\mu$ is numerically negligible, 
due to the partial cancellation between
the chargino and neutralino contributions and, more importantly, 
the absence of any $\tan\beta$ enhancement.

%
\subsection{${\widetilde \ell}_L {\widetilde B}{\widetilde W}$ models}
%
%
%
\begin{figure}[t!]
\centering
\includegraphics[scale=0.6]{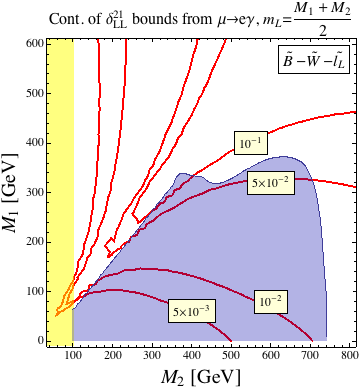}
\hspace{0.5cm}
\includegraphics[scale=0.6]{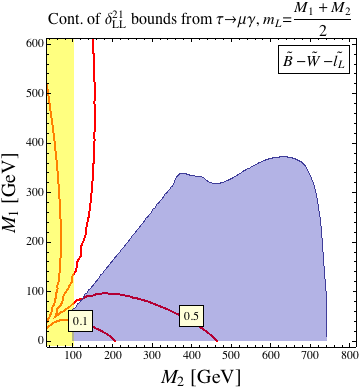}
\caption{Upper limits on $\delta^{21}_{LL}$ (left) and $\delta^{32}_{LL}$ (right) in the plane $(M_2,~M_1)$ for the 
model~$\widetilde{\ell}_L\widetilde{B}\widetilde W$ assuming $m_L = (M_1 + M_2)/2$. The light-blue area is excluded by the 
CMS~\cite{Khachatryan:2014qwa} direct search, the yellow region refers to the LEP exclusion.}
\label{fig:model_G}
\end{figure}
These models combine all the superpartners considered so far: the left handed sleptons,
the charged and neutral Winos, and the Bino. 
The right handed sleptons as well as the Higgsinos are assumed to be heavy,
and we therefore neglect Bino-Wino mixing.
As a result, it is again straightforward to compare the results of LHC lepton-based searches
to CLFV constraints: the spectrum is completely specified by the left-handed slepton masses,
which with no flavor dependence are given by $m_L$, the common Wino mass $M_2$ (up to possible
small splittings), and the Bino mass $M_1$, and, given the absence of Higgsinos and 
LR slepton mixing, only diagrams with the chirality flip occurring on the external
legs contribute,
\bea
\label{eq:BWL}
A_L &=& (A^{n_1}_{L})_{\mysmall U(1)} + (A^{n_1}_{L})_{\mysmall SU(2)} + (A^{c_1}_{L})_{\mysmall SU(2)}\,,
\qquad\qquad\qquad
A_R \simeq 0\,,
\\
\Delta a_\mu &=& \left(\Delta a^{n_1}_{\mu}\right)^{\!\mysmall L}_{\!\mysmall U(1)} + \left(\Delta a^{n_1}_{\mu}\right)_{\!\mysmall SU(2)} + 
\left(\Delta a^{c_1}_{\mu}\right)_{\!\mysmall SU(2)}\,,
\qquad\qquad~
d_e \simeq 0\,,
\eea
with no $\mu$ or $\tan\beta$ dependence.

At the same time, 
this model features a rich chargino-neutralino sector, 
and it has been employed by the LHC collaborations for the interpretation of 
searches based on multi-leptons plus missing energy~\cite{Khachatryan:2014qwa,Aad:2014nua},
assuming a Bino LSP, and sleptons half-way between the Bino and Wino. 
The highest sensitivity is reached in the case of heavy neutralino-chargino associated 
production, followed by decays to the Bino LSP through
intermediate on-shell sneutrinos and sleptons.
This decay chain leads to three-lepton events, with two OSSF leptons. 
In Fig.~\ref{fig:model_G}, we plot the upper bounds on $\delta^{21}_{LL}$ (left) 
and $\delta^{32}_{LL}$ (right) in the plane $(M_2,~M_1)$ with the left-handed slepton mass 
taken at the value $m_L = (M_1 + M_2)/2$,
which maximizes the LHC reach in the three-leptons plus missing transverse momentum channel. 
As we can see, Wino-like neutralino/chargino
masses are excluded by CMS up to 700 GeV for 
LSP masses below roughly 300 GeV~\cite{Khachatryan:2014qwa}.
On the other hand, the CLFV constraints are relatively mild in this entire region: there
is essentially no bound from $\tau\to \mu \gamma$, and $\delta^{21}_{LL}$ of few to 10\% is
allowed for the highest masses probed by the CMS search.
These mild constraints are a consequence of a cancellation occurring between $U(1)$ and $SU(2)$ contributions in 
Eq.~(\ref{eq:BWL}), which feature opposite signs. 
In fact, for the value we chose for $m_L$, the two contributions exactly cancel
when $M_1\approx M_2$.

Finally, since the only chirality flip is on the external fermion leg(s), 
there is no contribution to the electron EDM. 
However, $\Delta a_\mu$ can be induced but only at negligible levels, 
as in the ${\widetilde \ell}_L {\widetilde W}$ model, because of partial cancellations
of chargino and neutralino contributions and because there are no 
$\tan\beta$-enhanced contributions.


\subsection{ $\widetilde{\ell}_R\widetilde B$ models}
%
\begin{figure}[t!]
\centering
\includegraphics[scale=0.6]{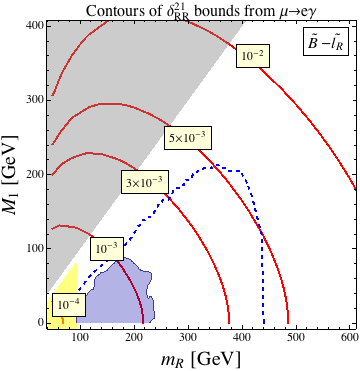}
\hspace{0.5cm}
\includegraphics[scale=0.6]{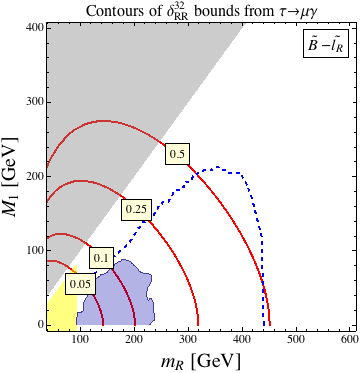}
\caption{
Upper limits on $\delta^{21}_{RR}$ (left) and $\delta^{32}_{RR}$ (right) in the plane  $(m_R,~M_1)$ for the 
model~$\widetilde{\ell}_R\widetilde B$. The light-blue area is excluded by the ATLAS~\cite{TheATLAScollaboration:2013hha} direct search, 
the yellow region refers to the LEP exclusion while in the grey area the LSP is not neutral. The dashed line refers
to the future LHC limit with $\sqrt{s}$=14 TeV and  $\mathcal{L}=100~\rm fb^{-1}$, as estimated in
\cite{Eckel:2014dza}.}
\label{fig:model_B}
\end{figure}
We now turn to models in which the light sleptons are right-handed, with the left-handed
sleptons decoupled.
The simplest of these contains just the Bino, in addition to the right handed sleptons.
Our discussion will be brief here, since it is essentially the same as the discussion
of the $\widetilde \ell_L\widetilde B$ model.

The simplified 
expressions for $A_{L,R}$, $\Delta a_\mu$, and $d_e$ can be obtained from~Eqs.~(\ref{eq:lLB}, \ref{eq:lLB2}),
with $L\leftrightarrow R$, and  $\alpha_Y\to 4\alpha_Y$, 
\beqa
A_R &=& (A^{n_1}_{R})_{\mysmall U(1)}\,,\qquad A_L \simeq 0\,,
\nonumber\\
\Delta a_\mu &=& \left(\Delta a^{n_1}_{\mu}\right)^{\!\mysmall R}_{\!\mysmall U(1)}\,, \qquad d_e 
\simeq 0\,.
\label{eq:1a}
\eeqa
As in the $\widetilde \ell_L\widetilde B$ model, the electron EDM $d_e$ vanishes, 
and $\Delta a_\mu$ is always negative and very small.
In Fig.~\ref{fig:model_B}, we show contours of the upper bounds on the dimensionless 
MI parameter 
$\delta^{21}_{RR} \equiv \Delta_{RR}^{21}/m^2_R$ (left) and $\delta^{32}_{RR} \equiv \Delta_{RR}^{32}/m^2_R$ (right) 
in the  $(m_R,~M_1)$ plane.

As already anticipated, the CLFV constraints are stronger than in the 
$\widetilde{\ell}_R\widetilde B$ model, because of the larger hypercharge of the 
right-handed sleptons. On the other hand, the cross-section for Drell-Yan production 
of left-handed slepton is larger than for right handed sleptons, resulting in a lower sensitivity
to the latter.

The discussion of this section carries over trivially to
${\widetilde \ell}_R {\widetilde B}{\widetilde W}$ models.
Since the right-handed
sleptons do not couple to pure Winos, the latter have no effect on either LHC 
slepton production or on
the dipole amplitudes.

Models with only right-handed sleptons and Winos are somewhat special, 
predicting, in particular, no dipole amplitudes.
We will briefly comment on such ``exotic'' models at the end of this Section.



\subsection{$\widetilde{\ell}_L\widetilde{\ell}_R\widetilde B$ models}
%
\begin{figure}[t!]
\centering
\includegraphics[scale=0.6]{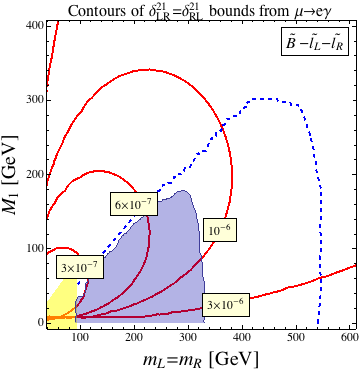}
\hspace{0.5cm}
\includegraphics[scale=0.6]{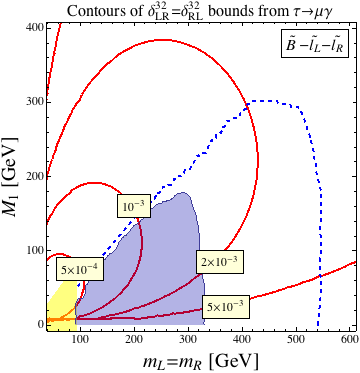}
\caption{
Upper limits on $\delta^{21}_{RL} = \delta^{21}_{LR}$ (left) and $\delta^{32}_{RL} = \delta^{32}_{LR}$ (right) in the plane 
$(m_L=m_R,~M_1)$ for the model~$\widetilde{\ell}_L\widetilde{\ell}_R\widetilde B$. The light-blue area is excluded by the 
ATLAS~\cite{TheATLAScollaboration:2013hha} direct search, the yellow region refers to the LEP exclusion. The dashed line refers
to the future LHC limit with $\sqrt{s}$=14 TeV and  $\mathcal{L}=100~\rm fb^{-1}$, as estimated in
\cite{Eckel:2014dza}.}
\label{fig:model_C3}
\end{figure}
%
%
\begin{figure}[t]
\centering
\includegraphics[scale=0.6]{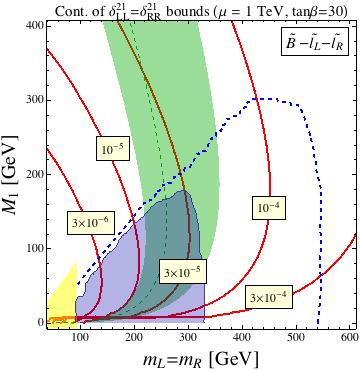}
\hspace{0.5cm}
\includegraphics[scale=0.6]{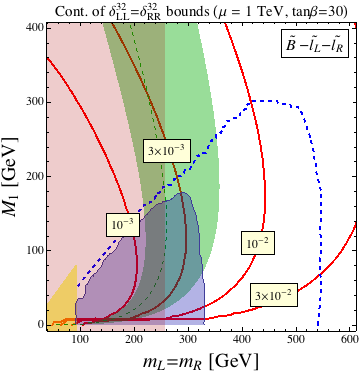}
\caption{
Upper limits on $\delta^{21}_{LL} = \delta^{21}_{RR}$ (left) and $\delta^{32}_{LL} = \delta^{32}_{RR}$ (right) in the plane 
$(m_L=m_R,~M_1)$ for $\widetilde{\ell}_L\widetilde{\ell}_R\widetilde B$ models 
assuming $\mu = 1$~TeV and $\tan\beta = 30$. 
The light-blue area is excluded by the ATLAS~\cite{TheATLAScollaboration:2013hha} direct search, the yellow region refers to the LEP exclusion. 
The dashed line refers to the future LHC limit with $\sqrt{s}$=14 TeV and  $\mathcal{L}=100~\rm fb^{-1}$, as estimated in
\cite{Eckel:2014dza}.
The green band accounts for the muon $g-2$ anomaly at the $2\sigma$ level: $\Delta a_\mu = (2.9\pm 1.8)\times 10^{-9}$.
The red-shaded area is excluded by stau sector constraints (see text for details). The constraints on the $\delta$'s scale as $(30~{\rm TeV})/({\mu\times\tan\beta})$.}
\label{fig:model_C1}
\end{figure}

Models with both left-handed and right-handed sleptons
are qualitatively different from the scenarios discussed above due
to the possibility of left-right mixing, which allows for a chirality flip on 
the slepton line, and therefore a significant enhancement of the 
dipole amplitudes.
As a result, analyzing the reach of LHC searches together with CLFV observables is
trickier in this case. 
Even a relatively small left-right mixing, which
has little effect on the slepton masses, and therefore on LHC observables,
can significantly alter the predictions for the CLFV transitions.
We also note that $d_e$ does not vanish here, since the relevant diagrams involves
the couplings of the two different fermion chiralities, and these are generically independent 
complex  numbers.
Finally, because of the enhancement mentioned above, 
large contribution to $(g-2)_\mu$ are possible, as we will see shortly.

The various dipole amplitudes are now given by,
\beqa
A_L &=& (A^{n_1}_{L})_{\mysmall U(1)} + (A^{n_3}_{L})_{\mysmall U(1)}\,,\qquad
A_R = (A^{n_1}_{R})_{\mysmall U(1)} + (A^{n_3}_{R})_{\mysmall U(1)}\,,
\label{eq:meg3a}\\
\Delta a_\mu &=& \left(\Delta a^{n_1}_{\mu}\right)^{\!\mysmall L}_{\!\mysmall U(1)} + 
\left(\Delta a^{n_1}_{\mu}\right)^{\!\mysmall R}_{\!\mysmall U(1)} + 
\left(\Delta a^{n_3}_{\mu}\right)_{\!\mysmall U(1)}\,,
\label{eq:da3a}\\
d_e &=& \left(d_e^{n_3}\right)_{\!\mysmall U(1)}\,.
\label{eq:de3a}
\eeqa
While the amplitudes $(A^{n_1}_{M})_{\mysmall U(1)}$, with $M=L,R$, 
involve only left handed or right handed sleptons, 
$(A^{n_3}_{M})_{\mysmall U(1)}$ are proportional to the left-right mixing,
which can in principle involve either same-generation, or different generation
sleptons. 

In Fig.~\ref{fig:model_C3}, we show the bounds on the MIs $\delta^{21}_{RL}=\delta^{21}_{LR}$ 
(left) and $\delta^{32}_{RL}=\delta^{32}_{LR}$ (right) in the plane $(m_L=m_R,~M_1)$. 
Note that we assume here degenerate left-handed and right-handed sleptons,
in order to allow for a direct comparison with LHC search results.
As before, the light-blue area highlights the exclusion set by the 
ATLAS analysis in~\cite{TheATLAScollaboration:2013hha} on smuon and selectron masses. 
As a reference, we also show the $\sqrt{s}= 14$ TeV LHC forecast as estimated
for the $\widetilde{\ell}_L\widetilde B$ model 
in~\cite{Eckel:2014dza}\footnote{Note however that this estimate assumes the presence of 
left-handed sleptons only.}.

We now turn to discuss contributions with
LL and RR flavor-violating insertions,
with the chirality flip coming from LR mixing of same-generation sleptons.
Treating the LR mixing as an insertion, we will again show our results in the  
$(m_L=m_R,~M_1)$ plane.
This is of course an approximation, since the LR mixing necessarily implies a splitting
of the two masses. However, as discussed above, even a small LR mixing, which has
little effect on the masses, can significantly alter the dipole operators.
In the following, we will also assume that the slepton $A$-terms are small,
so that the left-right slepton mixing is proportional to $\mu\tan\beta$.
The left-right mixings in the different generations are then correlated
(and proportional to the relevant lepton mass). Furthermore, they are 
also correlated with the Higgsino masses, which are $\sim\mu$ for large $\mu$.  
It is important to bear in mind however that these relations need not hold
generally, for example, if some A-terms are large, or if additional
parameters enter the Higgsino spectrum.

With these assumptions, the CLFV amplitudes, as well as $g-2$, are proportional to $|\mu\tan\beta|$.
Motivated by $g-2$, in Fig.~\ref{fig:model_C1} (left) we show the contours of 
 $\delta^{21}_{RR}$ and $\delta^{21}_{LL}$ for $\mu= 1$~TeV and $\tan\beta = 30$.
It is easy to reinterpret the CLFV bounds for different choices of these parameters:
roughly, the bounds on 
$\delta^{ij}_{RR}$ 
and $\delta^{ij}_{LL}$ scale 
as $(30~{\rm TeV})/({\mu\times\tan\beta})$.

Compared to the previously discussed 
models $\widetilde{\ell}_L\widetilde B$ and $\widetilde{\ell}_R\widetilde B$, 
we see that now, while the LHC bounds are only slightly more constraining, 
the bounds on the MIs are much more stringent. As a result, 
the present LHC direct exclusion is comparable with the limit from $\mu\to e\gamma$
only for values of the MIs
$\delta^{21}_{LL}=\delta^{21}_{RR} \lesssim 10^{-4}$, $\delta^{21}_{LR}=\delta^{21}_{RL} \lesssim 10^{-6}$.
Larger values of these LFV parameters would imply that the model is already excluded by MEG way beyond
the reach of the LHC.

As mentioned above, an interesting feature of these models 
is that they
can provide a supersymmetric contribution to $\Delta a_\mu$ with the right 
sign to reduce the tension 
between theoretical prediction and experiment. 
This is clearly seen in Fig.~\ref{fig:model_C1}, where the green band corresponds to
$\Delta a_\mu = (2.9\pm 1.8)\times 10^{-9}$, thus accounting for the $(g-2)_\mu$ anomaly at the $2\sigma$ level or better.

Similarly, the right panel of Fig.~\ref{fig:model_C1} displays the bounds
on  $\delta^{32}_{RR}$ and $\delta^{32}_{LL}$.
However, the interpretation of the parameters in the smuon-stau system is a bit more subtle.
Since the $LR$ slepton mixing is proportional to the lepton mass, a large $\mu\tanbeta$
can have a significant effect on the stau mass eigenvalues, lowering the mass of the 
lightest combination. 
Thus,  $m_L$ and $m_R$ stand for the smuon masses as before, while the stau masses
are in general different as a result of the LR mixing.
Furthermore,
for given values of $m_L$ and $m_R$, 
the LEP bound on the stau mass,
$m_{\widetilde{\tau}_1}\gtrsim 80$~GeV
implies an upper bound on  
$|\mu\tan\beta|$\footnote{LHC searches for direct EW production of staus have 
not reached the sensitivity  yet to set a stronger bound. See e.g.~\cite{ATLAStau}.}.
An even stronger constraint follows from the condition of (meta)stability of the vacuum,
since a large Higgs-stau-stau trilinear coupling in the potential can induce
a charge-breaking minimum  which is deeper than the correct electroweak breaking minimum.
This implies the following bound~\cite{Rattazzi:1996fb, Hisano:2010re, Sato:2012bf, Kitahara:2012pb}:
\begin{equation}
| \mu\times\tan\beta| \lesssim 39 (\sqrt{m_L}+\sqrt{m_R})^2- 10~{\rm TeV}.
\label{eq:vacuum}
\end{equation}
The resulting excluded area is shaded in red.

The portion of the plane favoured by $(g-2)_\mu$ can be easily enlarged to values of the SUSY masses above
the present and future LHC reach by increasing $|\mu\tanbeta|$. However, this would imply a much stronger constraint
on LFV in the $\mu-e$ sector, as well as a stau spectrum heavier than selectrons and smuons to overcome the meta-stability bound
of Eq.~(\ref{eq:vacuum}): this is shown in the right panel of Fig.~\ref{fig:model_C1}, where we see that for degenerate staus and sleptons
the stau sector constraints partly exclude the region favoured by $(g-2)_\mu$. 

Finally, let us comment on the electron EDM.
Clearly, for the large values of $g-2$ considered here,
an $\mathcal{O}(1)$ phase of $\mu M_1$ 
would be a phenomenological disaster. 
This is the well-known SUSY CP-problem:
the new experimental limit $d_e < 8.7\times 10^{-29}$~e~cm~\cite{Baron:2013eja}
requires $\arg(\mu M_1)$ at the level of $10^{-4}$ or less.

\subsection{$\widetilde{\ell}_{L}\widetilde{\ell}_{R}\widetilde W$ models}
In these scenarios, the Higgsinos and Bino are decoupled while the Winos, 
and left- and right-handed sleptons are light.
Since the Wino couples only to left-handed sleptons, the low-energy predictions of this model 
are the same as those of the previous simplified model $\widetilde{\ell}_L \widetilde W$, Eqs.~(\ref{eq:2b}, \ref{eq:amu2b}).
As in some previous cases, LHC searches have not been interpreted yet in this simplified scenario.
Nevertheless, given the presence of the left-handed sleptons, we expect them to be at least as stringent 
as in the case of model $\widetilde{\ell}_L \widetilde W$.

To summarize, the right-handed slepton plays a subdominant role and the phenomenology of this model is
also captured by Fig.~\ref{fig:model_D}.


%
\subsection{${\widetilde \ell}_L {\widetilde B}{\widetilde H}$ models}
%
%
\begin{figure}[t!]
\centering
\includegraphics[scale=0.6]{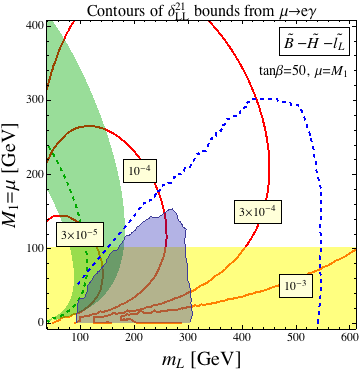}
\hspace{0.5cm}
\includegraphics[scale=0.6]{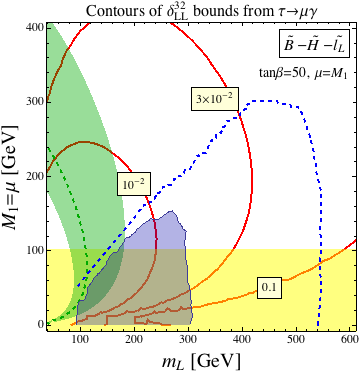}
\includegraphics[scale=0.6]{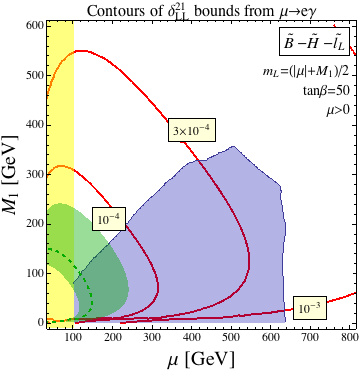}
\hspace{0.5cm}
\includegraphics[scale=0.6]{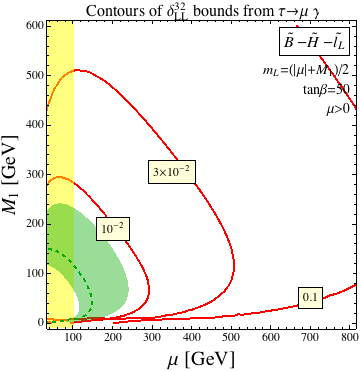}
\caption{Upper limits on $\delta^{21}_{LL}$ (left) and $\delta^{32}_{LL}$ (right) for the model~$\widetilde{\ell}_L\widetilde{B}\widetilde H$ 
in the  $(m_L,~M_1=\mu)$  plane (top panels) and in the
$(\mu,~M_1)$ plane    with $m_L = (M_1 + \mu)/2$ 
(bottom panels), for $\tan\beta=50$. 
The light-blue areas are excluded by ATLAS searches (see text for details), 
the yellow region shows the LEP exclusion. 
The dashed line refers to the future LHC limit with 
$\sqrt{s}$=14 TeV and  $\mathcal{L}=100~\rm fb^{-1}$, 
as estimated in \cite{Eckel:2014dza}.
The green band accounts for the muon $g-2$ anomaly at the $2\sigma$ 
level: $\Delta a_\mu = (2.9\pm 1.8)\times 10^{-9}$.
For lower $\tan\beta$, the constraints on the $\delta$'s weaken by a factor
50/$\tanbeta$.
}
\label{fig:model_H}
\end{figure}
We now turn to light Higgsino scenarios, starting with examples
in which the only light fields are the left-handed
sleptons, the Higgsinos and the Bino. The resulting amplitudes are:
\bea
A_L &=& (A^{n_1}_{L})_{\mysmall U(1)} + (A^{n_2}_{L})_{\mysmall U(1)}
\,,\qquad\qquad
A_R 
\simeq
0\,,
\\
\Delta a_\mu &=& \left(\Delta a^{n_1}_{\mu}\right)^{\!\mysmall L}_{\!\mysmall U(1)} + 
\left(\Delta a^{n_2}_{\mu}\right)^{\!\mysmall L}_{\!\mysmall U(1)}\,,
\qquad~~~
d_e = \left(d_e^{n_2}\right)^{\!\mysmall L}_{\!\mysmall U(1)}\,.
\eea
In addition to the
three scales $m_L$, $M_1$ and $\mu$, these amplitudes 
are sensitive to $\tan\beta$. 
The dominant contribution is typically from the Bino-Higgsino 
diagrams of $(A^{n_2}_{L})_{\mysmall U(1)}$, which are proportional to the
Bino-Higgsino mixing $\sim\mu\tan\beta$.
Thus,  for large $\tan\beta$, $\Delta a_\mu$ can account for 
the current anomaly  if  $\mu>0$.
In the following, we choose $\mu>0,~ \tan\beta=50$ to maximize 
$\Delta a_\mu$. 
Since the leading contributions to the $\ell_i\to \ell_j \gamma$ 
amplitudes scale as $\mu\tan\beta$, 
the CLFV bounds we derive below can easily be reinterpreted for lower values
of $\tan\beta$. Thus for example,  for $\tan\beta=5$, these bounds will weaken
by one order of magnitude.

As for the mass scales involved, we examine two benchmark scenarios. 
In the first, $M_1=\mu$, so that the light neutralino is a Bino-Higgsino
mixture.
In the second, $M_1$ and $\mu$ vary independently with $m_L = (M_1 + \mu)/2$,
in analogy with the previous models we considered.

In the top panels of Fig.~\ref{fig:model_H}, we plot
the upper bounds on $\delta^{21}_{LL}$ (left) and $\delta^{32}_{LL}$ (right) 
in the plane $(m_L,~M_1=\mu)$.
The green band highlights the region preferred by the muon $g-2$: 
$\Delta a_\mu = (2.9\pm 1.8)\times 10^{-9}$.
The ATLAS exclusion (light-blue area) and the future LHC prospects 
(dashed line) are the 
same as in the ${\widetilde \ell}_L {\widetilde B}$ models,
and are based on slepton pair production followed by their decay into
the (Bino component of the) neutralino LSP\footnote{ 
Further constraints from lepton based-searches could arise from sneutrino production followed by 
decays into leptons plus Higgsino-like charginos.}.
We also show in yellow the LEP exclusion on Higgsino-like charginos.
We see that the LFV bounds are quite stringent in this case.
In the region probed by the LHC lepton-based searches,
$\delta^{21}_{LL} < 10^{-4}$
and $\delta^{32}_{LL} < 10^{-2}$.

In the bottom panels of Fig.~\ref{fig:model_H}, we show the results 
for the second slice of the parameter space we chose:
independent $\mu$ and $M_1$ with intermediate sleptons, $m_L = (M_1 + \mu)/2$.
LHC searches can only probe these models for $M_1$ and $\mu$ which are 
sufficiently different, with the LSP being either a Bino or a neutral Higgsino
depending on the hierarchy of $M_1$ and $\mu$. 
While the low-energy constraints are similar to the previous case, 
the most sensitive searches at the LHC are based on Higgsino-like 
neutralino pair production. The subsequent decay
of the Higgsinos into the intermediate sleptons induces events with 4 leptons and missing energy.
Notice that such a decay preferably occur through the small gaugino components of the heavier neutralinos, hence
it is still democratic for the $e$ and $\mu$ flavors. 
On the other hand,
the Higgsinos would prefer to decay into staus, 
if kinematically accessible, especially for large $\tan\beta$.
Therefore, if the only light sleptons are the selectron and the smuon,
we can use the 4-lepton ATLAS search \cite{Aad:2014iza} 
to constrain these models
(see bottom left panel of Fig.~\ref{fig:model_H}). 
However, if the stau is light too,
the branching fractions of Higgsino decays to selectrons and smuons
will be very small. Such scenarios are still constrained by the LEP bound
(see bottom right panel of Fig.~\ref{fig:model_H}).
Additional constraints can be extracted from LHC searches for associate production of charginos and neutralinos decaying to intermediate staus, 
leading to events with three taus and missing energy~\cite{ATLAStau}.
For a light LSP, this search can set a limit on the Higgsino mass up to 350 GeV \cite{Calibbi:2013poa}.

Finally, notice that this model features a non-vanishing contribution to $d_e$. As discussed above for the model $\widetilde{\ell}_L\widetilde{\ell}_R\widetilde B$, present bounds then require a certain suppression of the flavor-blind phase $\arg(\mu M_1)$.

%
\subsection{${\widetilde \ell}_R {\widetilde B}{\widetilde H}$ models}
\begin{figure}[t]
\centering
\includegraphics[scale=0.6]{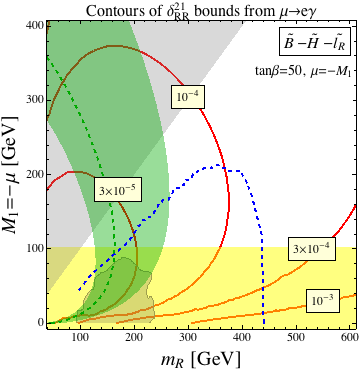}
\hspace{0.5cm}
\includegraphics[scale=0.6]{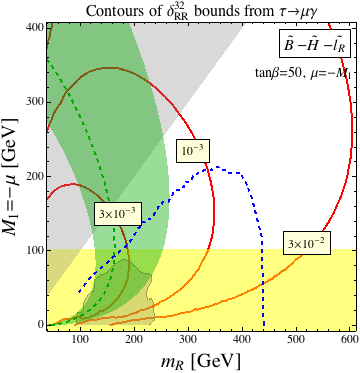}
\includegraphics[scale=0.6]{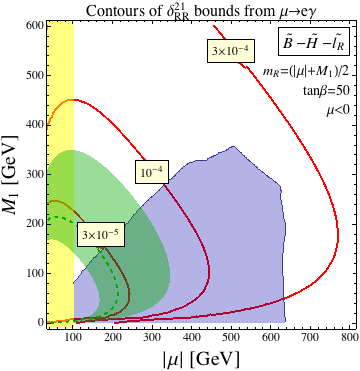}
\hspace{0.5cm}
\includegraphics[scale=0.6]{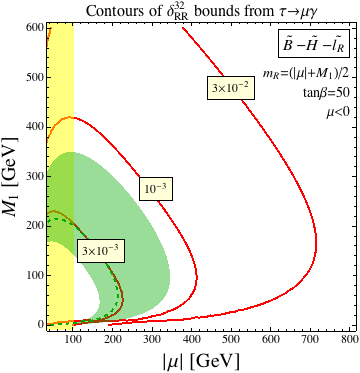}
\caption{Upper limits on $\delta^{21}_{RR}$ (left) and $\delta^{32}_{RR}$ (right) 
for $\widetilde{\ell}_R\widetilde{B}\widetilde H$ models in the
$(m_R,~M_1=-\mu)$  plane 
(top panels) and in the $(\mu,~M_1)$  plane with $m_R = (M_1 + |\mu|)/2$ 
(bottom panels), for 
$\tan\beta=50$. 
is assumed.
The light-blue areas are excluded by ATLAS searches (see the text for details), 
the yellow region refers to the LEP exclusion. The dashed line refers
to the future LHC limit with $\sqrt{s}$=14 TeV and  $\mathcal{L}=100~\rm fb^{-1}$, as estimated in
\cite{Eckel:2014dza}.
The green band accounts for the muon $g-2$ anomaly at the $2\sigma$ level: $\Delta a_\mu = (2.9\pm 1.8)\times 10^{-9}$.
For lower $\tan\beta$, the constraints on the $\delta$'s weaken by a factor
50/$\tanbeta$.
}
\label{fig:model_I}
\end{figure}
In this scenario, the left-handed sleptons and Winos are heavy while 
the right-handed
sleptons, the Higgsinos and the Bino are light. 
The amplitudes can be obtained by exchanging $R\to L$ in the previous
model,
\bea
A_R &=& (A^{n_1}_{R})_{\mysmall U(1)} + (A^{n_2}_{R})_{\mysmall U(1)}
\,,\qquad\qquad
A_L 
\simeq 0\,,
\\
\Delta a_\mu &=& \left(\Delta a^{n_1}_{\mu}\right)^{\!\mysmall R}_{\!\mysmall U(1)} + 
\left(\Delta a^{n_2}_{\mu}\right)^{\!\mysmall R}_{\!\mysmall U(1)}\,,
\qquad~~~
d_e = \left(d_e^{n_2}\right)^{\!\mysmall R}_{\!\mysmall U(1)}\,.
\eea
However, in this case,  $\mu<0$ is required in order to account for the muon $g-2$ anomaly.
As before, 
we maximize these contributions by choosing $\tan\beta=50$, 
keeping in mind that the CLFV bounds scale as $\tan\beta$.
The resulting bounds on $\delta^{21}_{RR}$ (left) and $\delta^{32}_{RR}$ (right)
 are shown in Fig.~\ref{fig:model_I} 
for two benchmark scenarios in complete analogy to the
${\widetilde \ell}_L {\widetilde B}{\widetilde H}$   models: 
 $(m_R,~M_1=-\mu)$ (top panels) and $(\mu,~M_1)$ with $m_R = (M_1 + \mu)/2$ 
(bottom panels).
Note that the low-energy bounds are somewhat stronger than those of 
Fig.~\ref{fig:model_H},
because of the larger hypercharge of the right-handed sleptons. 
The discussion of the LHC searches for 
${\widetilde \ell}_L {\widetilde B}{\widetilde H}$ 
carries over to this case as well.
We only note 
that in the top-panel plots of Fig.~\ref{fig:model_H} we employed the
 same search as in 
Fig.~\ref{fig:model_I}
and, in this case, we do not expect significant constraints
from other searches, given the absence of sneutrinos.

%
\subsection{${\widetilde \ell}_L {\widetilde W}{\widetilde H}$ models}
%
\begin{figure}[t]
\centering
\includegraphics[scale=0.6]{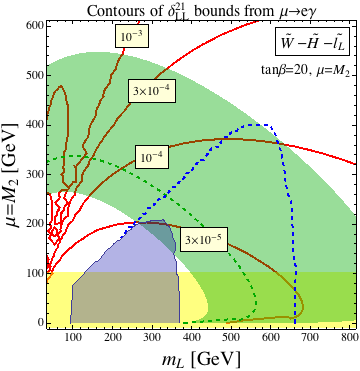}
\hspace{0.5cm}
\includegraphics[scale=0.6]{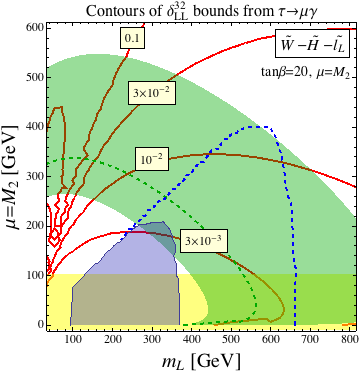}
\caption{Upper limits on $\delta^{21}_{LL}$ (left) and $\delta^{32}_{LL}$ (right) 
for $\widetilde{\ell}_L\widetilde{W}\widetilde H$ models, in the $(m_L,~M_2=\mu)$ 
plane for $\tan\beta=20$. 
The light-blue area (dashed line) represents the current (future) LHC  exclusion, 
with the latter taken from~\cite{Eckel:2014dza}.
The yellow region refers to the LEP exclusion.
The green band accounts for the muon $g-2$ anomaly at the $2\sigma$ level: $\Delta a_\mu = (2.9\pm 1.8)\times 10^{-9}$. The bounds on the $\delta$'s scale with $\tanbeta$.
}
\label{fig:model_J}
\end{figure}
The last class of models we discuss in detail has
left-handed
sleptons, Higgsinos and Winos. 
Only the right-handed sleptons and Bino are decoupled. 
The particle content of these models is rich, with three neutralinos, 
two charginos, as well as charged and neutral sleptons. 
Several diagrams therefore contribute to the dipole amplitudes, 
\bea
A_L &=& (A^{n_1}_{L})_{\mysmall SU(2)} +  (A^{c_1}_{L})_{\mysmall SU(2)} + (A^{n_2}_{L})_{\mysmall SU(2)}
+ (A^{c_2}_{L})_{\mysmall SU(2)}
\,,\qquad
A_R 
\simeq 0\,,
\\
\Delta a_\mu &=& \left(\Delta a^{n_1}_{\mu}\right)^{\!\mysmall L}_{\!\mysmall SU(2)} + \left(\Delta a^{c_1}_{\mu}\right)^{\!\mysmall L}_{\!\mysmall SU(2)} + 
\left(\Delta a^{n_2}_{\mu}\right)^{\!\mysmall L}_{\!\mysmall SU(2)} + \left(\Delta a^{c_2}_{\mu}\right)^{\!\mysmall L}_{\!\mysmall SU(2)}\,,
d_e = \left(d_e^{c_2}\right)_{\!\mysmall SU(2)}+\left(d_e^{n_2}\right)_{\!\mysmall SU(2)}\,.
\eea
These are typically dominated by Wino-Higgsino diagrams, which are $\tanbeta$ enhanced.
In particular, 
a large contribution to the muon $g-2$, 
from $\left(\Delta a^{c_2}_{\mu}\right)^{\!\mysmall L}_{\!\mysmall SU(2)}$,
is a pretty generic prediction of these scenarios. 
This can be clearly seen in Fig.~\ref{fig:model_J}, 
where we again
plot 
the bounds on $\delta^{21}_{LL}$ (left) and $\delta^{32}_{LL}$ (right),
for the simplifying choice 
$\mu=M_2$. The green band, corresponding to $\Delta a_\mu = (2.9\pm 1.8)\times 10^{-9}$, is particularly wide,
even for 
$\tan\beta=20$. The LHC exclusion is the same as in  
${\widetilde \ell}_L {\widetilde W}$ models, 
since DY-produced selectrons and smuons still prefer 
decaying into a Wino-like neutralino.
LFV constraints are also quite strong, again because of the $\tanbeta$ enhancement
of Wino-Higgsino diagrams,
although some cancellations are possible for $\mu=M_2 \approx 3 m_L$.

\subsection{$\widetilde{\ell}_R\widetilde W$ models}
Models in which only the right-handed sleptons and the Winos are light are somewhat special. 
Of course, if the Higgsinos, Bino, and left-handed sleptons were completely decoupled, 
the dipole transitions would vanish, and the right-handed sleptons would 
be long-lived. Realistically however, the leading contributions to dipole transitions arise from Wino/Bino mixing effects since the 
Bino couples to right-handed fields. In our setup, such mixings are roughly given by $\sim (m^2_Z t^{-1}_{\beta})/(\mu M_1)$ 
and therefore very suppressed. As a result, all low-energy observables receive negligible effects.

As for collider searches, we note that the right-handed sleptons will still decay to the LSP, through the 
small Bino component. If such a mixing is suppressed enough, the decay could occur at a displaced vertex or even outside the detector.
However, this would require super-heavy Bino and Higgsinos.
In this case the relevant bound would come from charged track searches, as quoted at the beginning of the section.
If on the contrary the decay is prompt, the constraints from direct LHC searches should resemble those of model $\widetilde{\ell}_R \widetilde B$. 
On the other hand, a Wino-like lightest neutralino corresponds to an almost degenerate chargino, and thus one has to take 
into account a lower bound on the spectrum from chargino searches at LEP: 
$M_2\gtrsim 100$ GeV. 


\subsection{Models with no light gauginos: ${\widetilde \ell}_L {\widetilde H}$, ${\widetilde \ell}_R {\widetilde H}$, ${\widetilde \ell}_L{\widetilde \ell}_R {\widetilde H}$}
Models without light gauginos, as well as models with no Bino and only right-handed sleptons (like the previously discussed 
${\widetilde \ell}_R {\widetilde W}$ and its possible extension ${\widetilde \ell}_R {\widetilde W} {\widetilde H}$), are of little interest for our 
discussion of the interplay between LFV observables and collider searches, since the low-energy processes are suppressed to negligible 
rates by small couplings of the Higgsinos to the sleptons.
Nevertheless, they can have an interesting LHC phenomenology, as mentioned above for the ${\widetilde \ell}_R {\widetilde W}$ case.
Further examples are provided in  \cite{Eckel:2014dza}, where it is shown, for instance, that models like  ${\widetilde \ell}_L {\widetilde H}$
can be constrained more strongly than  ${\widetilde \ell}_L {\widetilde W}$, 
for certain choices of the parameters.
These scenarios  are better probed at collider experiments, 
with low-energy observables providing little sensitivity.

\subsection{Heavy superpartner decoupling}
%
Our results for the different simplified models can be taken at face value:
we have used CLFV searches to constrain new particles with the
quantum numbers of charged sleptons, gauginos and Higgsinos.
Naturally however, in order to interpret these results in the context of supersymmetry,
one must estimate the effects of the heavy superpartners which we omitted.
This is especially relevant for the CLFV constraints, which generically fall 
off as the second power of the superpartner scale, while LHC cross sections fall much more steeply.

By comparing the different examples above we can get a qualitative 
estimate for the importance of different superpartners.
The largest contributions involve either Higgsinos, or left-right slepton
mixing. The former depends on $\mu\tanbeta$, and decouple as $M_1\tan\beta/\mu$
for large $\mu$.
If Higgsinos are decoupled, and in the absence of LR slepton mixing,
bounds on $\delta_{LL}$ are hardly affected by right-handed sleptons and vice-versa.
This is the case in the first four models we discussed.
Comparing the 
${\widetilde \ell}_L {\widetilde B}{\widetilde W}$
 models to the $\widetilde{\ell}_L\widetilde B$
or $\widetilde{\ell}_L\widetilde W$ models we can see that
the effects of heavier Binos or Winos are small.
As discussed above, the results are much more sensitive to heavier Higgsinos.
Similarly, if the light sleptons are predominantly left handed, but with a
small admixture of right handed sleptons, the CLFV constraints are sensitive to the
heavier slepton states. 
%
\begin{table}[t]
\centering
\renewcommand{\arraystretch}{1.25}
\setlength{\extrarowheight}{3pt}
\begin{tabular}{|c|c|}
\hline
model & region of validity  \\
\hline\hline
 $\widetilde{\ell}_L\widetilde B$ &   For~$M_1\lesssim 150$~GeV:
$m_R\gtrsim 2$~TeV (i.e.~$m_R/m_L \gtrsim 5\div10$), $M_2 \gtrsim 500$~GeV 
\\ 
$\ $ & and $\mu \gtrsim 1$~TeV (3~TeV if $\tan\beta\gtrsim 10$)
  \\ 
$\ $&  For $M_1\gtrsim 300$ GeV: $m_R\gtrsim 10$ TeV (i.e.~$m_R/m_L \gtrsim 25$), 
$\mu \gtrsim 4$~TeV
and $\tan\beta\lesssim 5$\\\cline{1-2}
$\widetilde{\ell}_L\widetilde W$ &  $M_2\lesssim 180$ GeV, $\tan\beta\lesssim 3$, $\mu \gtrsim 12$ TeV and \\
$\ $ &   $M_1\gtrsim 3$ TeV  or $m_R\gtrsim 2$ TeV \\ \cline{1-2}
$\widetilde{\ell}_{L}\widetilde{B}\widetilde W$&
$\mu \gtrsim 2$~TeV, $m_R\gtrsim 2$~TeV (i.e.~$m_R/m_L \gtrsim 5\div10$)
\\ \cline{1-2}
$\widetilde{\ell}_R\widetilde B$ &
$m_L\gtrsim 2$~TeV (i.e.~$m_L/m_R \gtrsim 5\div10$), $\mu \gtrsim 1$~TeV 
(2.5~TeV for $\tan\beta\gtrsim 10$) \\ \cline{1-2}
$\widetilde{\ell}_{L}\widetilde{\ell}_{R}\widetilde B$ &
$\mu\gtrsim 500$~GeV  \\ \cline{1-2}
${\widetilde \ell}_L {\widetilde B}{\widetilde H}$ &
$M_2 \gtrsim 2$~TeV  \\ \cline{1-2}
 ${\widetilde \ell}_R {\widetilde B}{\widetilde H}$ &
$m_L/m_R \gtrsim 2$  \\ \cline{1-2}
 ${\widetilde \ell}_L {\widetilde W}{\widetilde H}$  &
$m_R \gtrsim \mu/2$   \\ \cline{1-2}
\hline
\end{tabular}
\caption{Region of validity of CLFV estimates for the different 
simplified models.}
\label{tab:valid}
\end{table}

To estimate the importance of decoupled superpartners in each of the
models, we vary the parameters $m_L$, $m_R$, $M_1$, $M_2$, $\mu$ and 
$\tan\beta$,
and require that the CLFV amplitude used to derive the bounds above
is at least 5 times larger than all other amplitudes.
We note that this is a very strong requirement.
The largest amplitudes are always the $\mu\tan\beta$ enhanced ones,
coming from either Higgsino diagrams or from left-right slepton mixing.
With no $A$-terms, both these effects are controlled by $\mu\tan\beta$,
and cannot be disentangled. The conditions we find are collected
in Table~\ref{tab:valid}.

\section{Implications of LFV for LHC searches}\label{sec:SM_high}
We now turn to discuss the possible impact of slepton flavor dependence on 
different LHC searches. 
Specifically, we will only consider lepton plus missing energy searches in
simplified models containing sleptons, Binos and Winos.
We limit our discussion to models with sleptons of a single chirality and 
a neutralino LSP.
To simplify notation, we therefore
omit the chirality index of the sleptons.
The basic production process, common to all of these models,
is Drell-Yan slepton pair 
production, with each slepton decaying to one lepton and the LSP.

In some of the models, chargino-chargino,  chargino-neutralino, or neutralino-neutralino pair
production are possible too. 
These have a much higher reach compared to Drell-Yan production, because
of the larger cross-sections.
In the following we will discuss these different processes in turn.

Flavor-blind simplified models containing sleptons and neutralinos/charginos were analyzed 
by ATLAS and CMS.
Since our aim is to estimate the effects of flavor dependence on these searches,
we start by qualitatively reproducing the relevant exclusion for each of 
the flavor-blind models, and then repeat the analysis in the presence of some
slepton mass splitting and/or mixing.
We use CheckMATE~\cite{CheckMATE} to reinterpret the searches\footnote{
CheckMATE relies on several code packages and algorithms: the Delphes 3 detector 
simulation~\cite{Delphes}, the FastJet package \cite{Cacciari:2011ma, Cacciari:2005hq} which implements many sequential recombination algorithms (such as anti-$k_T$ \cite{AntiKt}), and the CLs prescription \cite{CLs} for statistical discrimination.}.
We therefore concentrate on several ATLAS analyses which are incorporated
and validated in CheckMATE.
Signal events are generated using MadGraph5\_aMC@NLO~\cite{MG5},
with the showering performed by the PYTHIA package~\cite{PYTHIA6}.

\subsection{$\widetilde{\ell}_L\widetilde B$, $\ell=e,\mu$ models}
In $\widetilde{\ell}_L\widetilde B$ models, sleptons are only produced
via $\gamma^*$- or $Z$-mediated Drell-Yan processes.
Since the slepton couplings to the photon and the $Z$ are  diagonal
in the slepton mass basis,
these processes result in 
$\widetilde \ell_i^+ \widetilde \ell_i^-$ pairs with $i=1,2$,
and flavor mixing has no effect on the production\footnote{
Note that flavor mixing could enter through  LR $Z$ coupling, but we neglect LR mixing throughout this Section.}.
On the other hand, flavor mixing has an important role in slepton decays.
In the presence of nonzero mixing, each slepton mass eigenstate can decay
to the LSP in association with either an electron or a muon,
so that slepton pair production leads to missing energy and $e^\pm\mu^\mp$
pairs, in addition to OSSF lepton pairs.
This may affect the sensitivity of searches based on OSSF leptons.
Furthermore,  Opposite Sign Different Flavor
(OSDF) dileptons, specifically $e^\pm\mu^\mp$ pairs, are sometimes used in
data-driven background estimates, with the assumption that the SUSY signal 
has no contribution in these channels.

\begin{figure}[h!]
\centering
\subfigure[~flavor blind: $\Delta m=0,~\sin2\theta=0$]{
\includegraphics[width=0.47\textwidth]{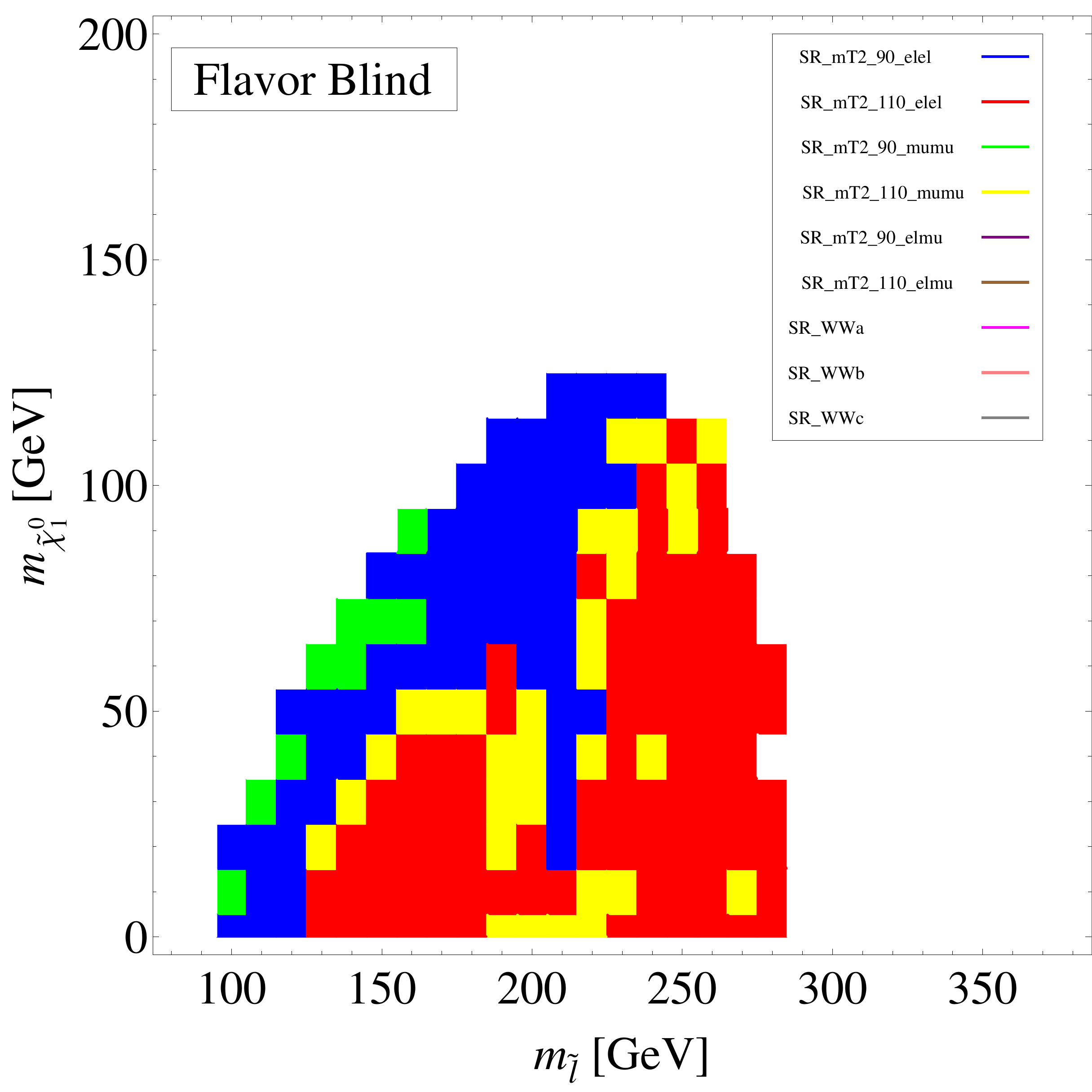}
\label{fig:DY_sleps_flavorblind}
}
\subfigure[~decoupled $\tilde\mu$, $\sin2\theta=0$]{
\includegraphics[width=0.47\textwidth]{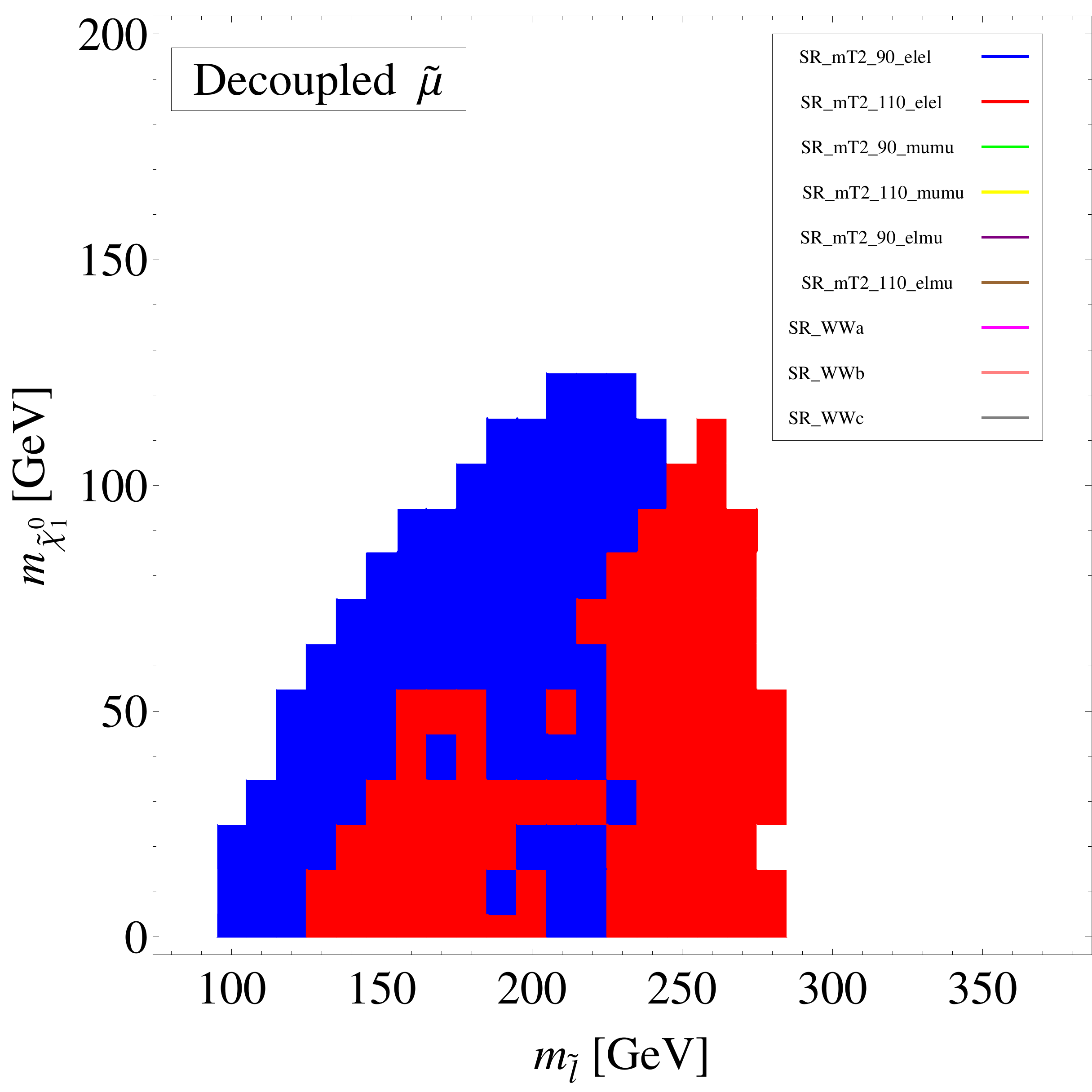}
\label{fig:DY_sleps_decoupled_smuon}
}
\\
\subfigure[~large mixing: small~$\Delta m$, $\sin2\theta=1$]{
\includegraphics[width=0.47\textwidth]{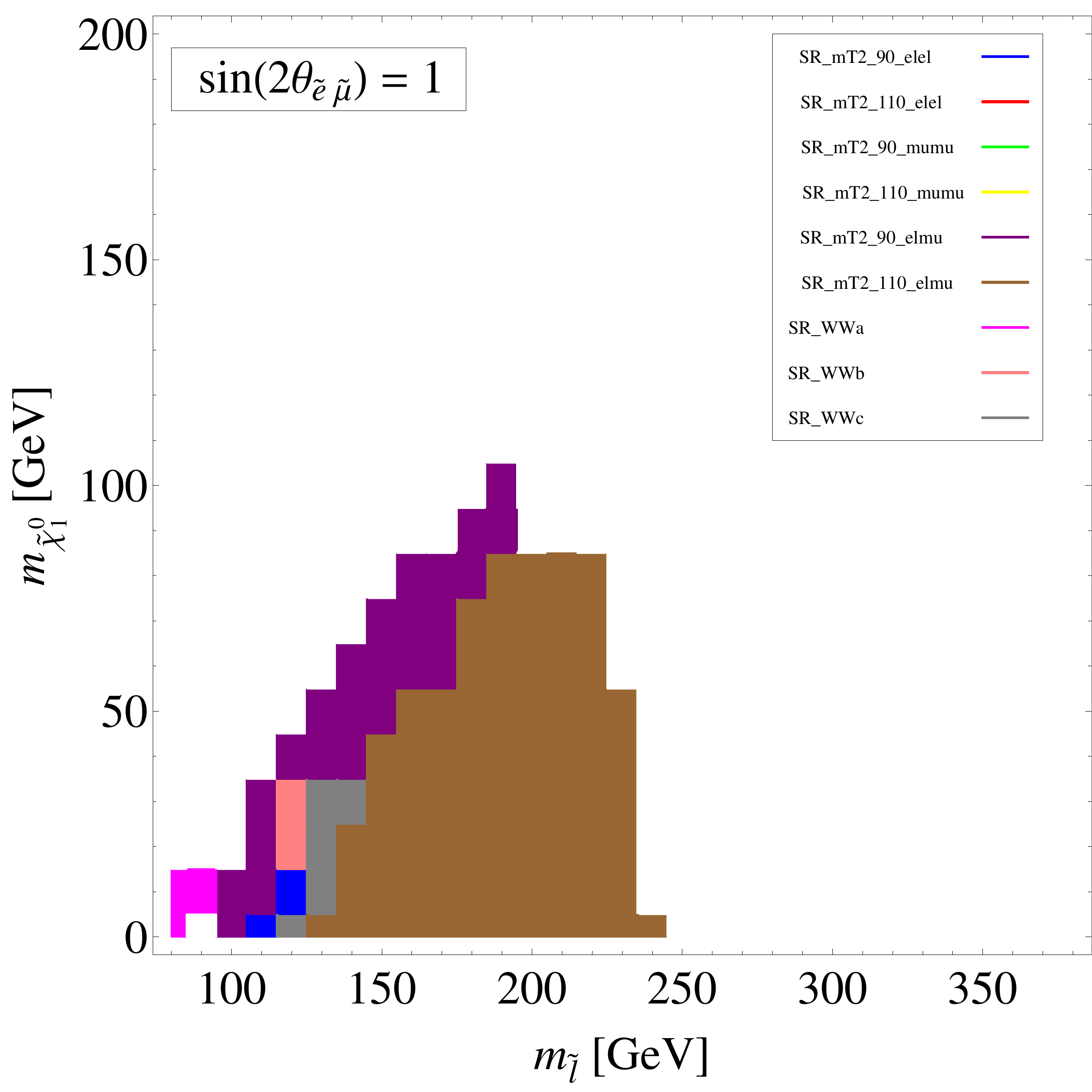}
\label{fig:DY_sleps_mixed_0p25}
}
\subfigure[~large mixing: small $\Delta m$, $\sin2\theta=0.8$]{
\includegraphics[width=0.47\textwidth]{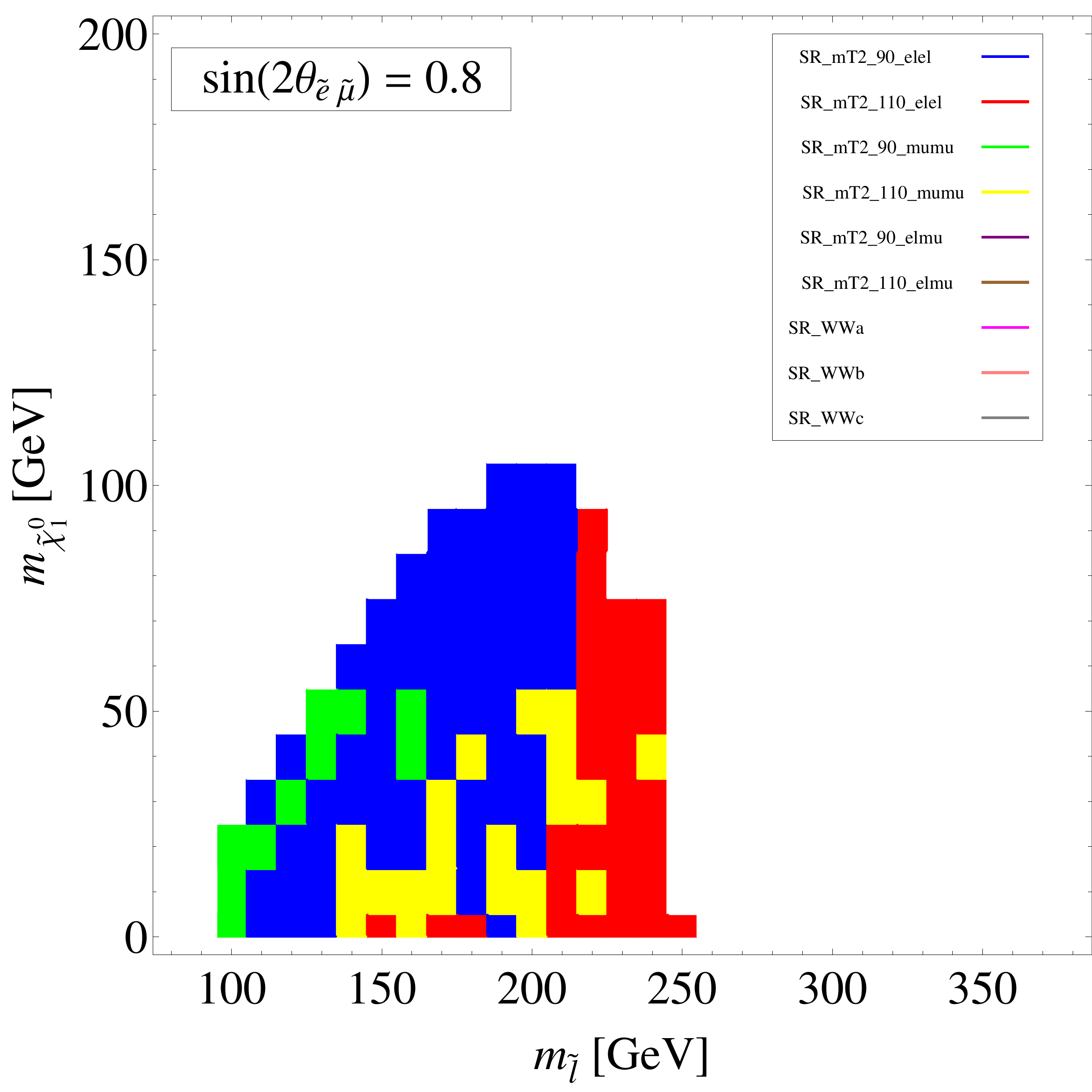}
\label{fig:DY_sleps_mixed_0p15}
}
\caption{
Reinterpreting the ATLAS analysis~\cite{TheATLAScollaboration:2013hha} 
to set limits on $\widetilde{\ell}_L\widetilde B$  models ($\ell=e,\mu$)
with different assumptions about flavor: (a) degenerate sleptons with no mixing;
(b) selectron only; (c), (d) almost degenerate sleptons with
 $\sin\theta=1$ and $\sin\theta=0.8$ respectively.
$m_{\tilde l}$ denotes the common slepton mass in (a), (c), (d),
and the selectron mass in (b).
The excluded region is color-coded according
to the most sensitive exclusion channel at each point (see legend). }
\label{fig:DY_sleps}
\end{figure}

We first reproduce the results of the ATLAS search for slepton pair production,
based on final states with 
OSSF dileptons plus missing 
energy~\cite{TheATLAScollaboration:2013hha}, assuming
degenerate selectrons and smuons with no 
flavor mixing\footnote{The more recent analysis~\cite{Aad:2014vma} 
has been embedded into CheckMATE but has not been validated. 
Nevertheless, the updated limits are very similar to those employed here.}.
We apply a flat K-factor of 1.3 to the leading-order cross-section 
as calculated by MadGraph5, in order to reproduce the cross-section
quoted in~\cite{TheATLAScollaboration:2013hha}.
The excluded region in the slepton-LSP mass plane is displayed 
in Fig.~\ref{fig:DY_sleps_flavorblind}. 
%
Note that the common slepton mass here coincides with $m_L$,
and the LSP mass is given by $M_1$.
The excluded region is color coded
to indicate the most sensitive exclusion channel at each point.
Thus for example, SR$_-$mT$_{2-}90_-$elel requires an electron-positron pair
with stranverse mass~\cite{Lester:1999tx, Barr:2003rg}
above 90~GeV. As expected, the different models are excluded  
by the $e^\pm e^\mp$ and $\mu^\pm \mu^\mp$ channels.
Note however, that the search~\cite{TheATLAScollaboration:2013hha} 
is sensitive to additional final states,
since it also targets chargino pair production followed either
by slepton-mediated chargino decays to leptons,
or by gaugino-mediated decays to $W$'s. The latter are important if the sleptons 
are heavy, and motivate SR$_-$WWa,  SR$_-$WWb and  SR$_-$WWc
(see legend of~Fig.~\ref{fig:DY_sleps_flavorblind})  which target $W$ pairs 
and missing energy.
Slepton-mediated chargino decays on the other hand lead to OS dileptons and missing
energy, with no correlation between the two lepton flavors.
These channels motivate SR$_-$mT$_{2-}90_-$elmu and  SR$_-$mT$_{2-}110_-$elmu,
which involve $e^\pm\mu^\mp$
and mT$_{2}$ above 90~GeV and 110~GeV respectively, and will be relevant for our discussion below.

We now  consider the possibility of flavor dependent
slepton masses.
As a first estimate  of the allowed flavor parameters, 
we start from the low-energy
bounds on $\delta^{21}_{LL}$ derived in the previous section using the 
MI approximation. 
Examining Fig.~\ref{fig:model_A}, 
we see that in the relevant region of the parameter space, 
the allowed values of $\delta_{LL}^{21}$ vary between $10^{-3}-10^{-2}$.
We can then translate these into allowed regions in the slepton masses 
and mixing. 
As noted above, the MI approximation fails for large relative mass splittings,
so throughout this section we use the full expressions reported in the appendix 
to obtain the CLFV constraints on the slepton 
parameters.

In the limit of 
small mixings and large mass splittings,
the cross-sections for selectron pair production and smuon pair production
can be very different.
Furthermore, since the efficiency of the search decreases as the slepton
mass approaches the LSP mass, large slepton mass differences
would result in different efficiencies for selectron and smuon discovery.
The LHC signatures of such models are  essentially the same 
however as in flavor-blind
models: $e^+e^-$ plus missing energy, and $\mu^+\mu^-$ plus missing energy.
Thus, in this limit, the ATLAS analysis, which treats the $ee$ and $\mu\mu$
samples separately, does more than place bounds
on degenerate selectrons and smuons. 
Rather, it separately constrains the selectron mass
and smuon mass.
Indeed, this flavor information is displayed in the updated 
ATLAS analysis,
which exhibits the separate limits on the selectron and smuon 
in the auxiliary plots~\cite{atlasaux}.

For completeness, we illustrate this point by fixing the smuon mass at 400~GeV, 
well above the lower bounds of Fig.~\ref{fig:DY_sleps_flavorblind}.
We then use CheckMATE to reinterpret the ATLAS 
search~\cite{TheATLAScollaboration:2013hha}
for models with different  selectron and Bino masses.
The results are shown in Fig.~\ref{fig:DY_sleps_decoupled_smuon}.
%
As expected, the exclusion limit for the selectron remains virtually unchanged.

In the opposite limiting case, the sleptons are almost degenerate,
and large mixings are allowed.
The production cross-sections of the two slepton
mass eigenstates are  practically equal. 
Thus, signal events redistribute among the $ee$, $\mu\mu$ and $e\mu$ final states
with fractions
\beqa
\label{eq:event_ratios}
\frac{N(e^\pm \mu^\mp)}{N(e^+e^-)} &=& 
\frac{ \sin^2 2\theta}{\cos^4\theta + \sin^4 \theta} \,,\nn
\\
N(\mu^+ \mu^-)&\sim&N(e^+e^-)\,.
\label{eq:signal_fracs}
\eeqa
Typically, searches based on just OSSF dileptons lose sensitivity in this scenario,
with signal events ``leaking'' into $e^\pm\mu^\mp$ final states.
However, as discussed above, the analysis of~\cite{TheATLAScollaboration:2013hha} 
is sensitive to $e^\pm\mu^\mp$ final states too.
The modified limits obtained for maximal mixing, $\sin2\theta=1$,
and for a mixing of $\sin2\theta=0.8$, are shown respectively in 
Fig.~\ref{fig:DY_sleps_mixed_0p25}
and Fig.~\ref{fig:DY_sleps_mixed_0p15}, assuming almost degenerate sleptons.
Indeed, for the maximal mixing case the most sensitive exclusion channels
are those involving $e\mu$ pairs, while for $\sin2\theta=0.8$
the OSSF channels are the dominant ones.
Either way, the reach in the slepton mass
is reduced by roughly 50~GeV compared to the flavor blind models,
and the reach in the Bino mass goes down by about 40~GeV.

\begin{figure}[h!]
\centering
\subfigure[~$\frac{\Delta m}{m}=5\times10^{-3},~\sin2\theta=1$]{
\includegraphics[width=0.47\textwidth]{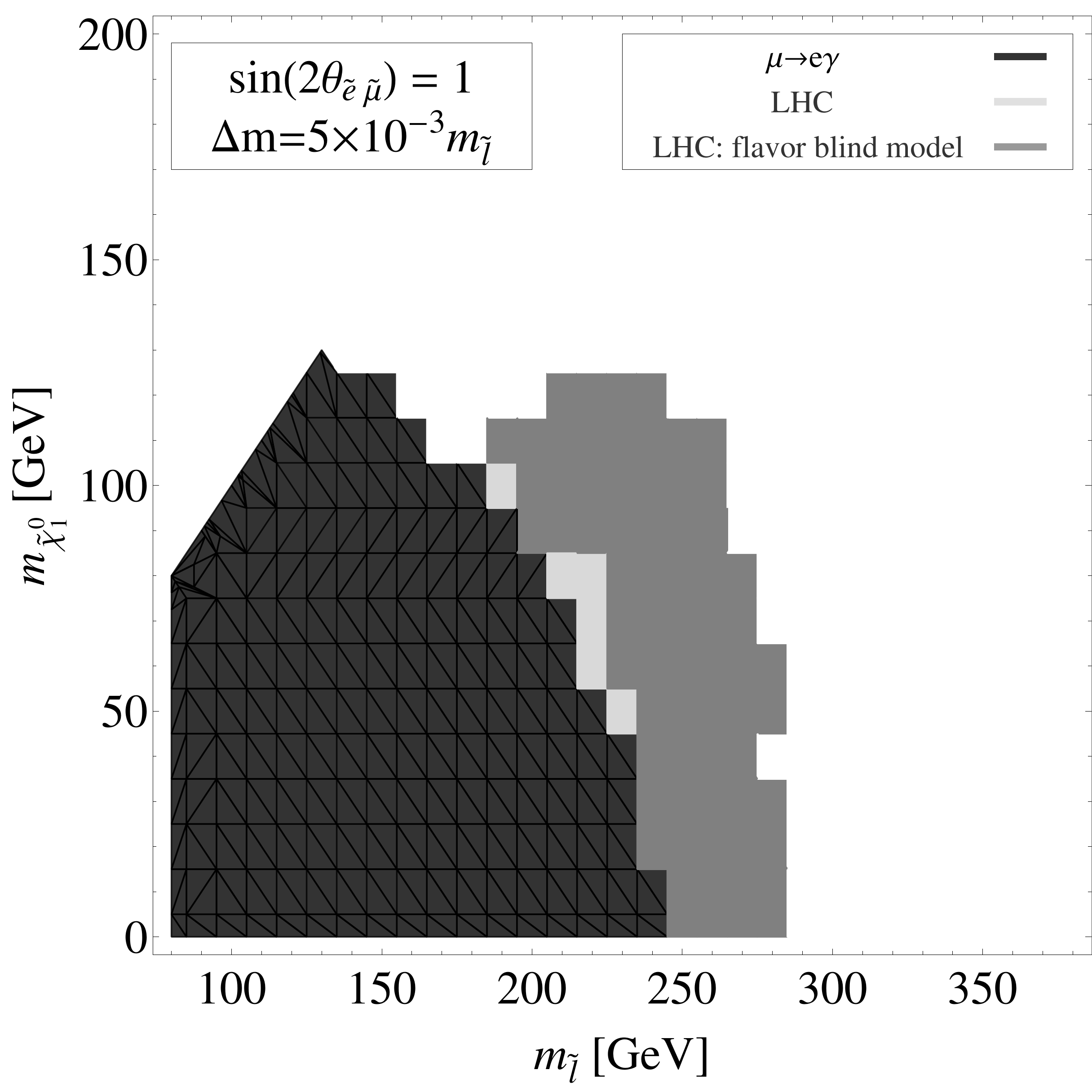}
\label{fig:DY_sleps_overlayed_constraints_5sin1}
}
\subfigure[~$\frac{\Delta m}{m}=5\times10^{-3},~\sin2\theta=0.8$]{
\includegraphics[width=0.47\textwidth]{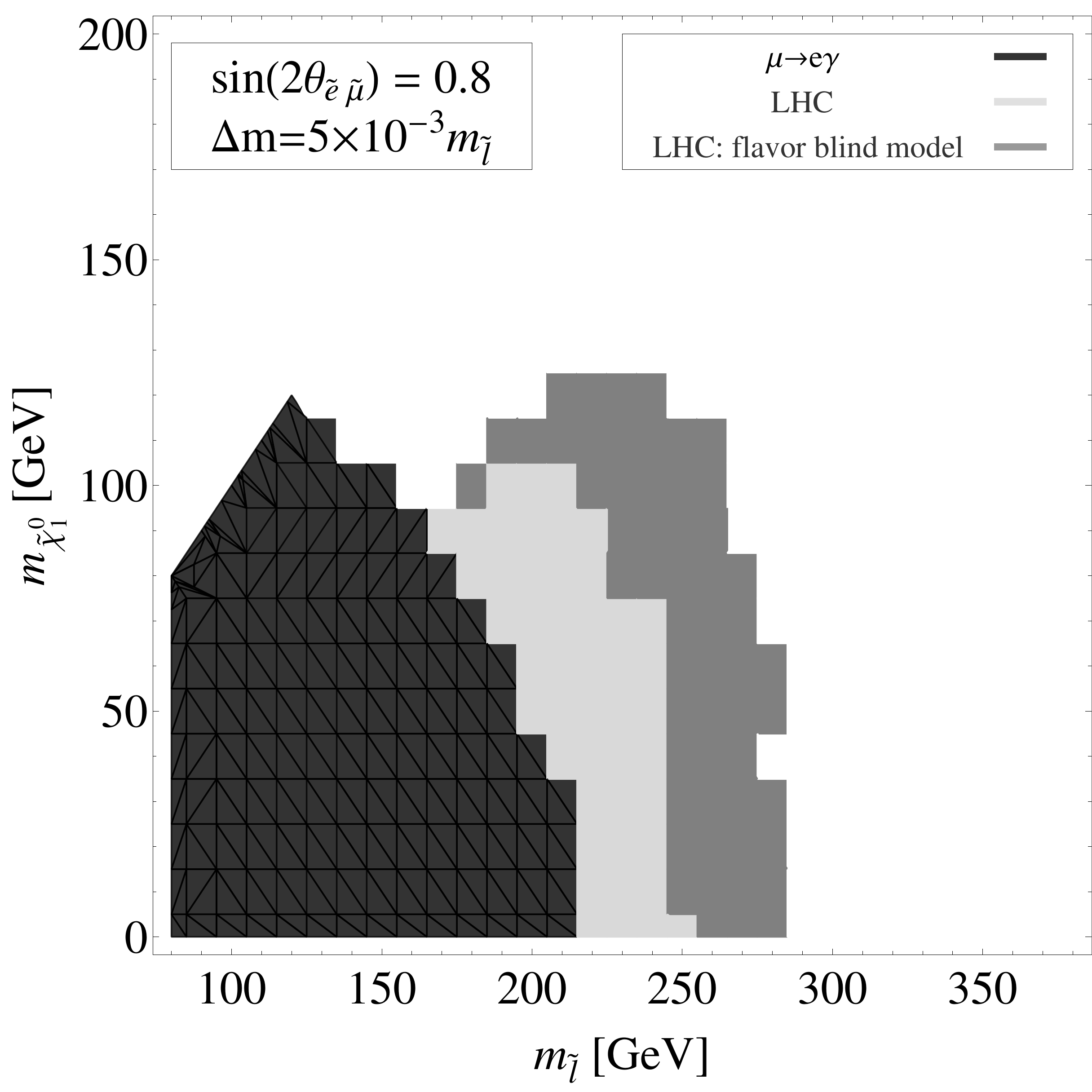}
\label{fig:DY_sleps_overlayed_constraints_5sinp8}
}
\\
\subfigure[~$\frac{\Delta m}{m}=3\times10^{-3},~\sin2\theta=1$]{
\includegraphics[width=0.47\textwidth]{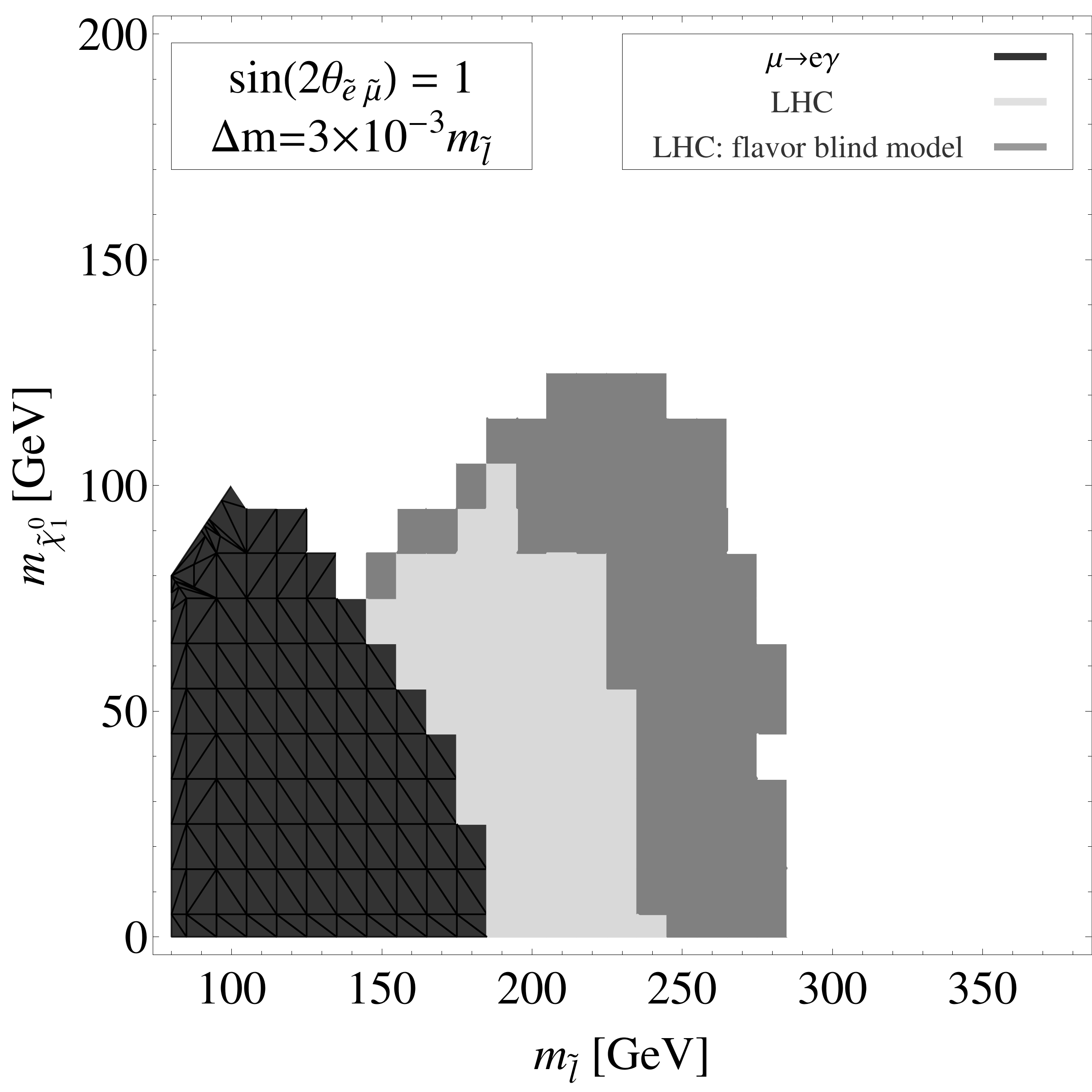}
\label{fig:DY_sleps_overlayed_constraints_3sin1}
}
\subfigure[~$\frac{\Delta m}{m}=3\times10^{-3},~\sin2\theta=0.8$]{
\includegraphics[width=0.47\textwidth]{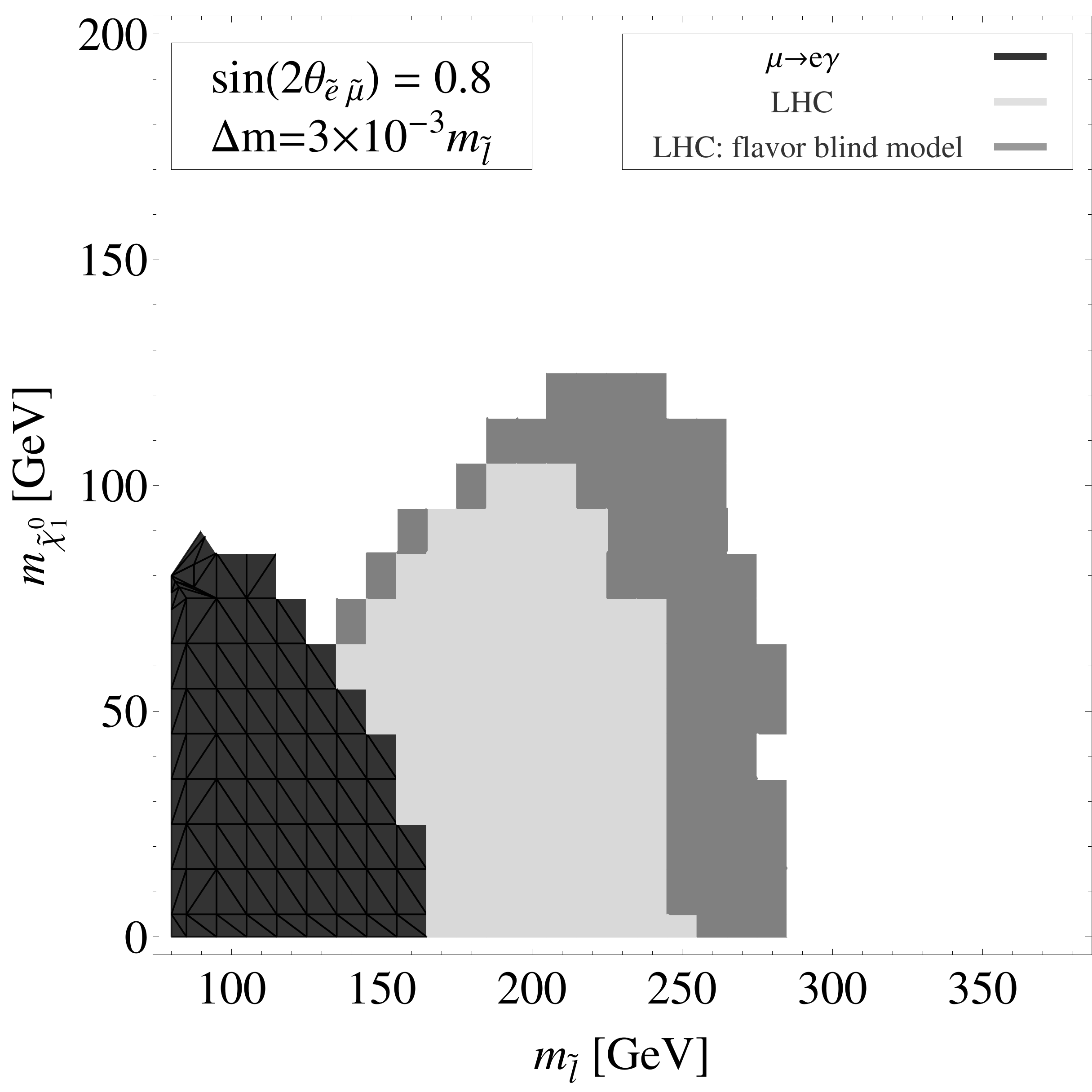}
\label{fig:DY_sleps_overlayed_constraints_3sinp8}
}
\caption{
The excluded region (light grey), in the $m_{\widetilde \ell} - m_\chi$ plane, 
obtained by reinterpreting the ATLAS analysis 
for almost degenerate sleptons (selectron-smuon) 
with relative mass splittings $\Delta m/m=5\times 10^{-3}$ (top) and
$\Delta m/m=3\times 10^{-3}$ (bottom),
for maximal mixing (left)
 and for $\sin2\theta=0.8$ (right).
The dark grey indicates the excluded region of 
Fig.~\ref{fig:DY_sleps_flavorblind} (flavor blind sleptons)
which is now allowed.
The dark hatched region is excluded by  $\mu\to e\gamma$.
}
\label{fig:DY_sleps_overlayed_constraints}
\end{figure}

Naturally, some of the parameter space displayed in 
Fig.~\ref{fig:DY_sleps_mixed_0p25}
and Fig.~\ref{fig:DY_sleps_mixed_0p15} is excluded by $\mu\to e\gamma$.
In Fig~\ref{fig:DY_sleps_overlayed_constraints} 
we show this constraint 
(dark hatched region), for two values of the slepton mass splitting, 
$\Delta m= 5\times10^{-3}\, m_{\tilde \ell}$ (upper panels) and 
$\Delta m = 3\times10^{-3}\, m_{\tilde \ell}$ (lower panels). 
We also reproduce here the
region excluded by the ATLAS data (light grey),
corresponding to  
Figures~\ref{fig:DY_sleps_mixed_0p25},\ref{fig:DY_sleps_mixed_0p15},
as well as 
the original excluded region (dark grey) of the flavor-blind model
as in Fig.~\ref{fig:DY_sleps_flavorblind}.
%
%

For mass splittings of order the slepton widths, slepton flavor
oscillations may be important~\cite{ArkaniHamed:1996au,ArkaniHamed:1997km}, and
as the mass splitting becomes much smaller than the width,
the fraction of $e^\pm \mu^\mp$ final states tends to zero.
Specifically, the cross-section for slepton pair production followed by
their decay to a final state with $e^\pm \mu^\mp$ and missing energy is
given by~\cite{ArkaniHamed:1997km},
\beq\sigma_{e\mu}^{\text{pair}} = \sigma_{0}^{\text{pair}} 
\frac{\sin^2 2\theta}{2}~r_\Gamma 
\label{eq:sigma-mue}
\eeq
where $\sigma_0^{\text{pair}}$ denotes the cross section in the absence 
of flavor mixing, 
and $r_\Gamma$ encodes the finite width effects, 
\beq
r_\Gamma \equiv \frac{3x^2 + x^4}{(1 + x^2)^2}
\eeq
with $x\equiv \Delta m/\Gamma$.
We verified that for the parameters of 
Fig.~\ref{fig:DY_sleps_overlayed_constraints},
$r_\Gamma\sim1$ so that the finite width effects are very small.

We see that for  $\Delta m= 5\times10^{-3}\, m_{\tilde \ell}$ with maximal mixing,
LHC searches and $\mu\to e\gamma$ have comparable sensitivity to the models,
while for all other choices, with smaller values of $\sin2\theta \Delta m/m$,
the LHC has better sensitivity.
Finally, the reduced LHC sensitivity compared to the flavor-blind case is
clearly seen in these plots.

\subsection{$\widetilde{\ell}_R\widetilde B$, $\ell=e,\mu$ models}
The discussion of the previous section carries over to this case as well,
but the allowed flavor effects are milder. 
As can be seen in Fig.~\ref{fig:model_B}, the allowed $\delta^{12}_{RR}$
is at most one or two permille throughout the parameter space probed by 
current searches.

For very small mixings and large mass differences, the ATLAS search yields separate bounds
on the R selectron and smuon.
For order-one mixings, the relative mass splittings has to be at 
the permille level because of the bound on $\delta^{12}_{RR}$.
However, the R-sleptons width is $\Gamma\lesssim 0.0045\times m_R$. 
Thus, the fraction of $e\mu$ final states is damped by the small
$r_\Gamma$. 
Flavor mixing effects will be  relevant however at the 14~TeV LHC, since 
for the higher mass scales probed at 14~TeV,
mass splittings larger than 
$\Gamma$ are compatible with $\mu\to e\gamma$ (see Fig.~\ref{fig:model_B}).

\subsection{$\widetilde{\ell}_L\widetilde B$, $\ell=\mu,\tau$ models}
As can be seen in Fig.~\ref{fig:model_A}, very large flavor effects are possible
in this case. We again distinguish between two limiting cases.
With small stau-smuon mixing, $\mu^\pm\mu^\mp$ pairs plus missing energy
have the same sensitivity to the smuon as in the flavor-blind scenarios.
Large smuon-stau mixings on the other hand, lead to
a smaller branching ratio for $\mu^\pm\mu^\mp$,
with some slepton pairs decaying to opposite sign muons and taus,
which largely escape detection (except possibly when the tau decays to a muon).
However, the selectron in this case is constrained to be a pure state,
and, if it is close in mass to the smuon, the searches are still sensitive
to the selectron through $e^+e^-$ plus missing energy channels.

\subsection{${\widetilde \ell}_L {\widetilde B}{\widetilde W}$ models}
\subsubsection{Limits from ${\widetilde\chi}^+{\widetilde\chi}^-$ production}
Here the signature of interest is two opposite sign
leptons plus missing energy,
coming from chargino pair production, with each chargino
decaying into a charged lepton, a neutrino, and the LSP, via either
a slepton or a sneutrino. 
These channels were used to set limits on the models in~\cite{TheATLAScollaboration:2013hha},
assuming six degenerate left-handed sleptons,
$\widetilde{e}$, $\widetilde\mu$, $\widetilde\tau$ plus three sneutrinos, 
with mass halfway between the chargino and neutralino.
The chargino was assumed to be 95\% Wino with a 5\% Higgsino component,
and the LSP a pure Bino\footnote{Note that the only difference between the  ${\widetilde \ell}_L {\widetilde B}{\widetilde W}$ models
considered here and the simplified models considered in~\cite{TheATLAScollaboration:2013hha},
is the small Higgsino component in the heavier neutralino and charginos.
This has little  effect however for left handed sleptons,
and particularly for the smuon and selectron.
}.  

\begin{figure}[h]
\centering
\subfigure[~flavor blind: $\Delta m=0$, no mixing]{
\includegraphics[width=0.47\textwidth]{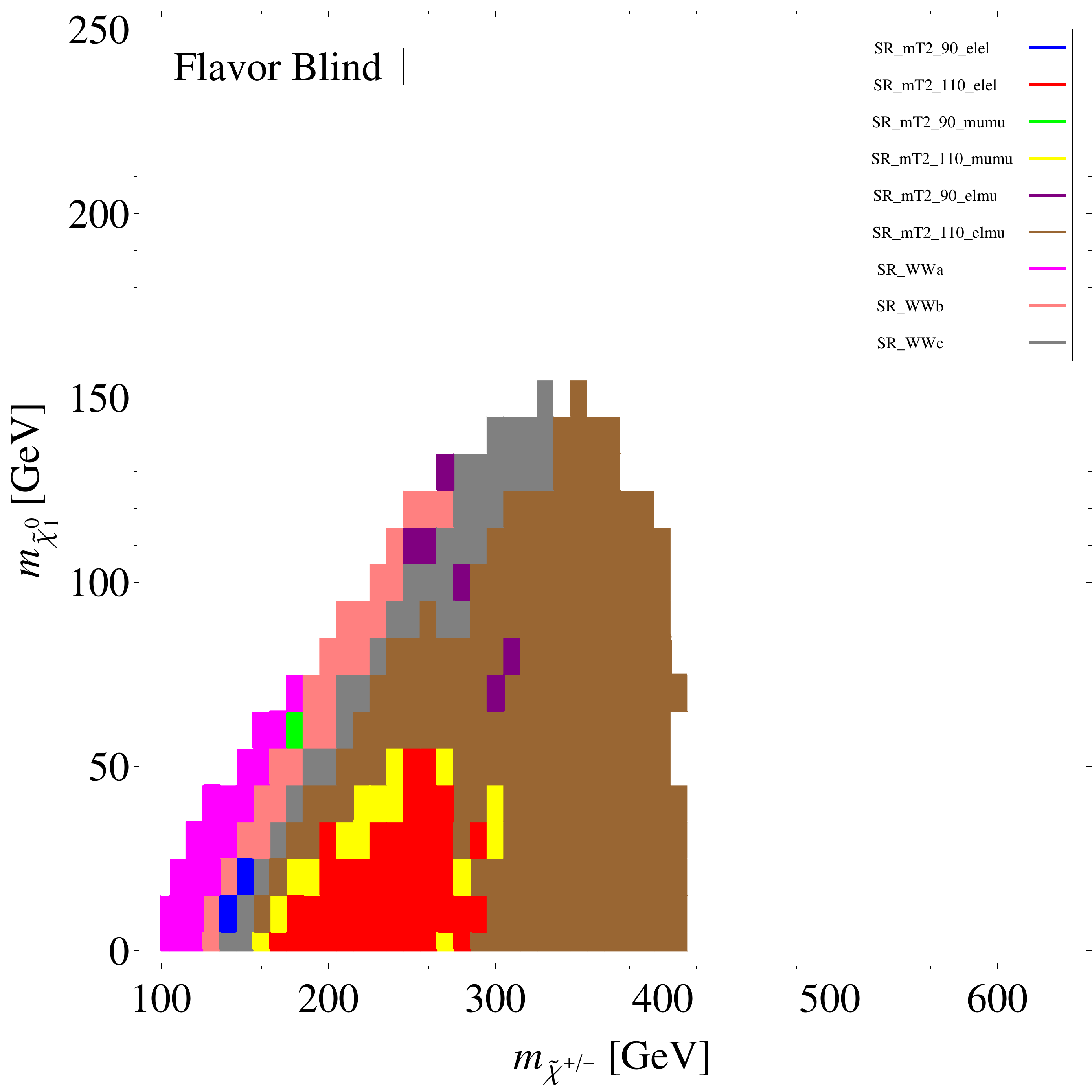}
\label{fig:chichi_flavorblind}
}
%
%
\subfigure[~ $\tilde\mu,\tilde\nu_\mu$ decoupled at 600~GeV, no mixing]{
\includegraphics[width=0.47\textwidth]{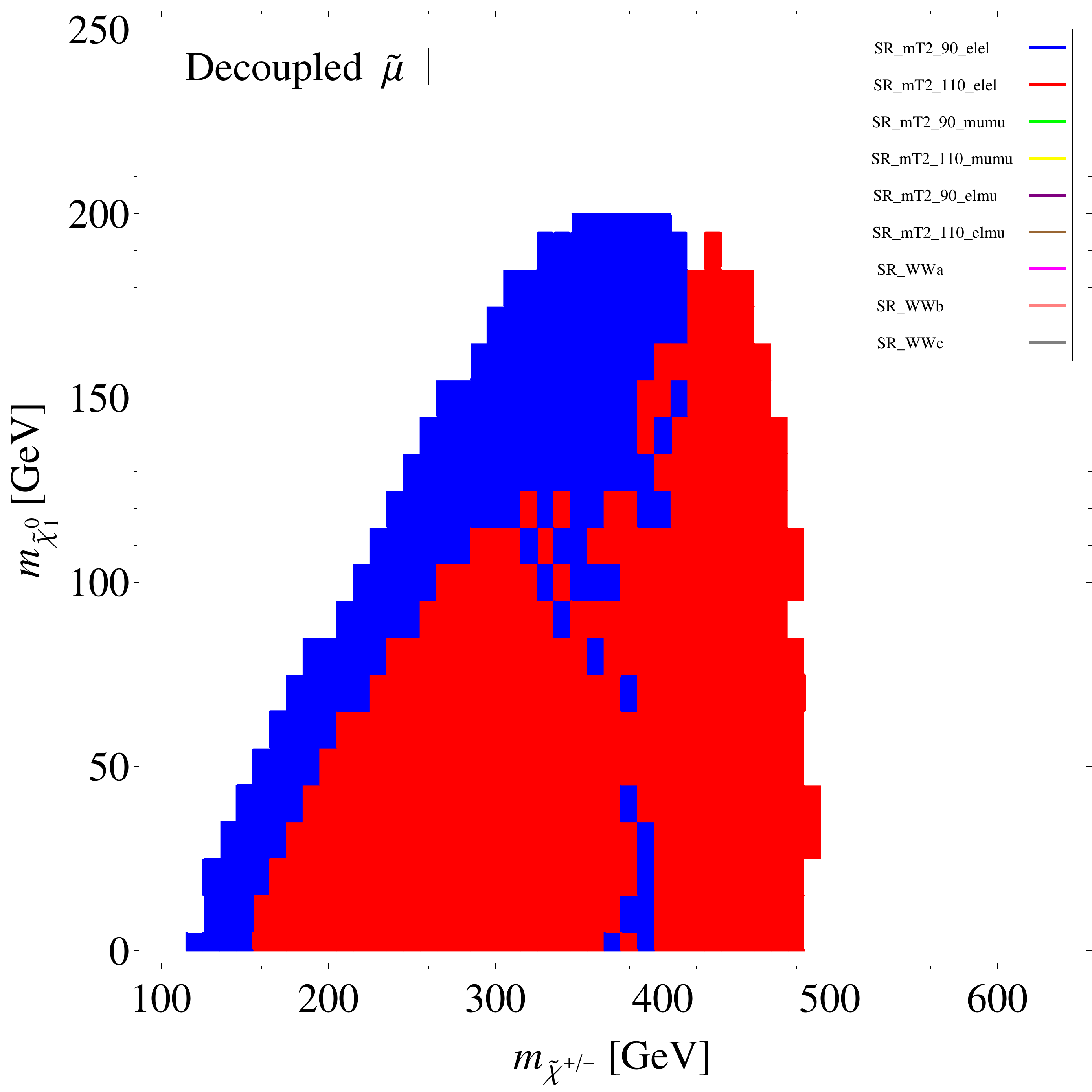}
\label{fig:chichi_split}
}
\caption{
The excluded region in the chargino-LSP mass plane, 
obtained by reinterpreting the search for chargino pair production~\cite{TheATLAScollaboration:2013hha} with subsequesnt decays to OS dileptons plus missing energy
for different flavor assumptions:
(a) six degenerate L-sleptons halfway between the chargino and LSP masses, 
with no flavor mixing; (b) same as in (a) but with the smuon and muon-sneutrino
decoupled at 600 GeV.
The color-coding is as in Fig.~\ref{fig:DY_sleps}.
}
\label{fig:chichi}
\end{figure}

Since the two leptons originate from
different charginos their flavors are not correlated, and flavor mixing has
no effect on this search.
However, a smuon-selectron mass difference can actually {\sl improve} the reach
for the light slepton in this case.
The reason, again, is related to the fact that the 
analysis~\cite{TheATLAScollaboration:2013hha} 
utilizes information from the
different lepton channels: $e^+e^-$, $e^\pm\mu^\mp$, etc.
With six degenerate sleptons, a chargino decays to either a charged slepton or a sneutrino
with equal probability, so the branching fraction for chargino decay to an electron
(plus invisible particles) is 1/3. If however, the smuon and the muon-sneutrino
are much heavier, this branching fraction goes up to 1/2.
In Figs.~\ref{fig:chichi_flavorblind} and~\ref{fig:chichi_split},
we compare the limits on the flavor blind models,
to the limits on the same models with the smuon decoupled at 600~GeV.
%
Indeed, the reach for the chargino mass is increased by about 100~GeV, while
the sensitivity to the LSP mass increases roughly from 150~GeV to 200~GeV. 
While we chose a large smuon mass for simplicity, even a much smaller mass
difference between the selectron and smuon would have an effect.

\subsubsection{Limits from ${\widetilde\chi}^\pm{\widetilde\chi}^0_2$ production}
\begin{figure}[h!]
\centering
\subfigure[~flavor blind: $\Delta m=0$, no mixing]{
\includegraphics[width=0.47\textwidth]{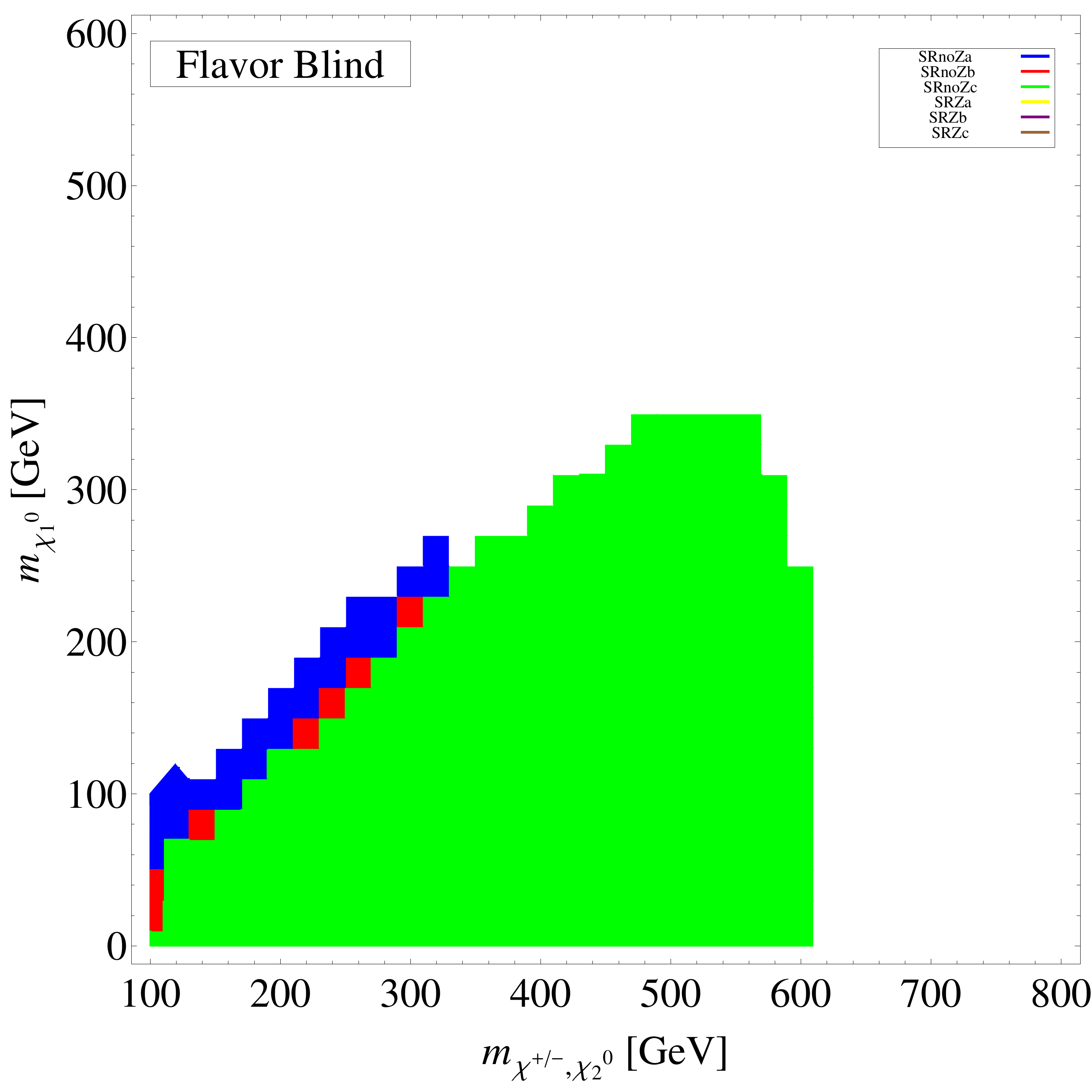}
\label{fig:neut_chi_flavorblind}
}
\subfigure[~mixed: small $\Delta m$, $\sin2\theta_{e\mu}=1$]{
\includegraphics[width=0.47\textwidth]{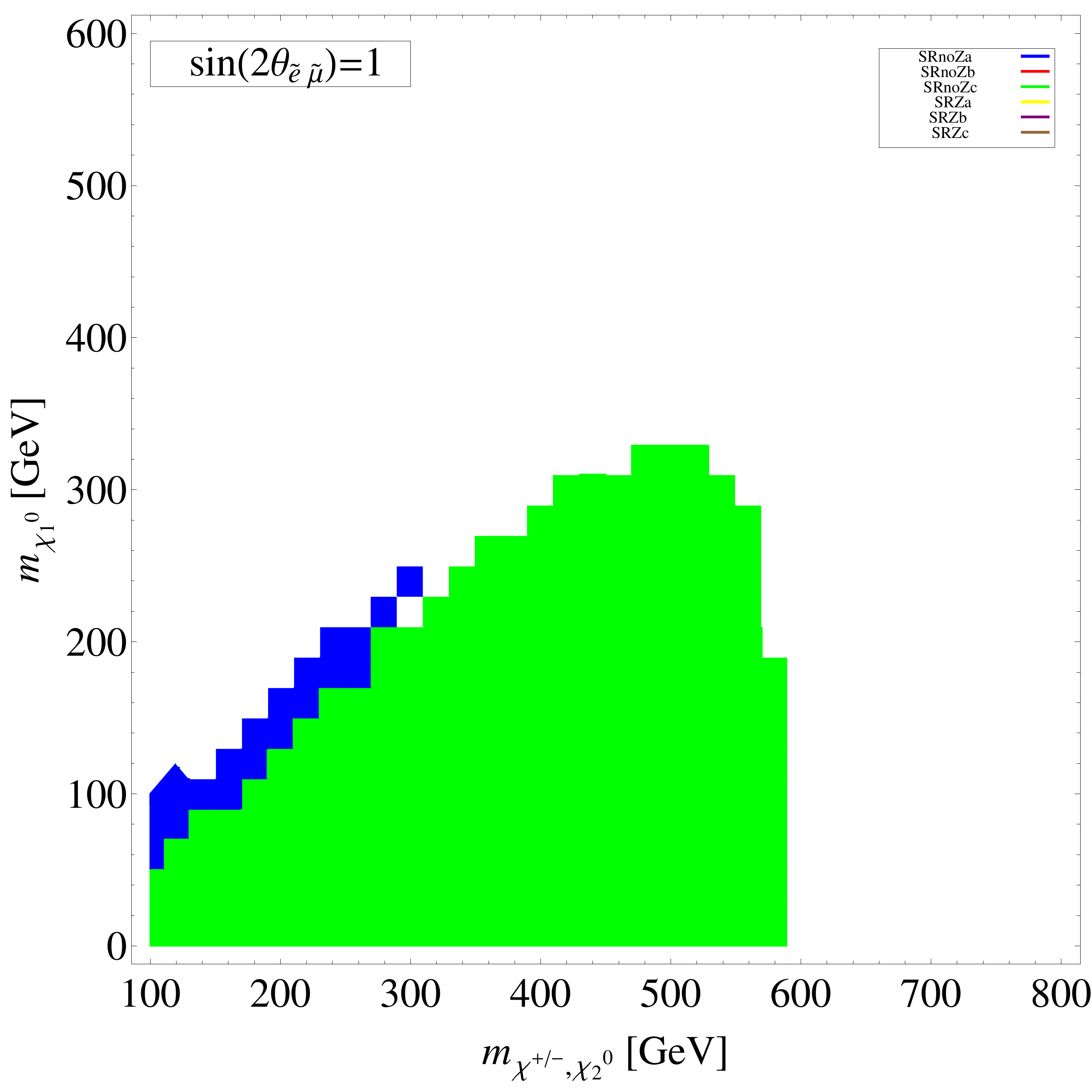}
\label{fig:neutchi_large_mixing}
}
\\
\subfigure[~decoupled $\tilde\mu, \tilde\nu_{\mu}$, no mixing]{
\includegraphics[width=0.47\textwidth]{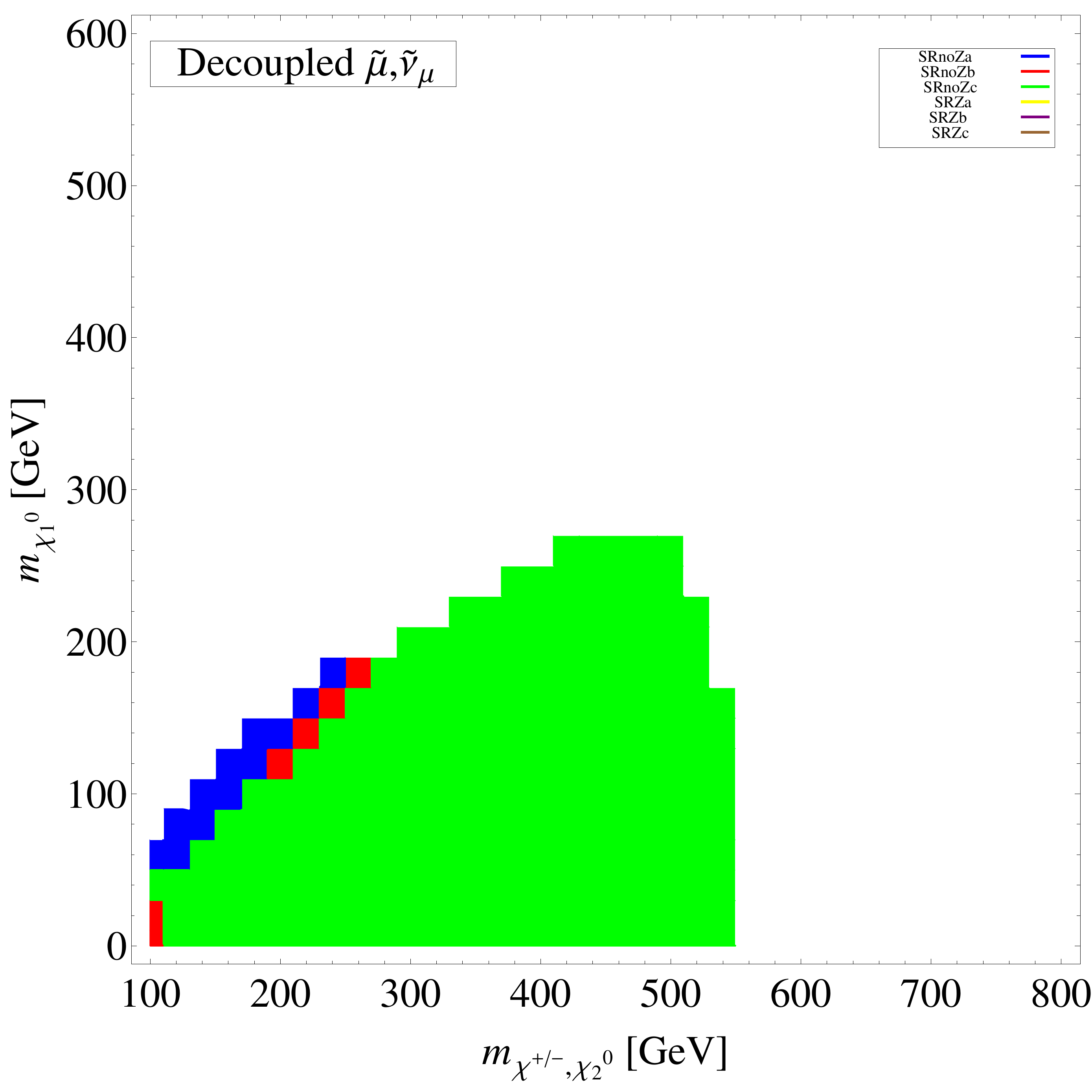}
\label{fig:neut_chi_split}
}
\subfigure[~decoupled $\tilde e, \tilde\nu_e$, $\sin2\theta_{\tilde\mu\tilde\tau}=1$]{
\includegraphics[width=0.47\textwidth]{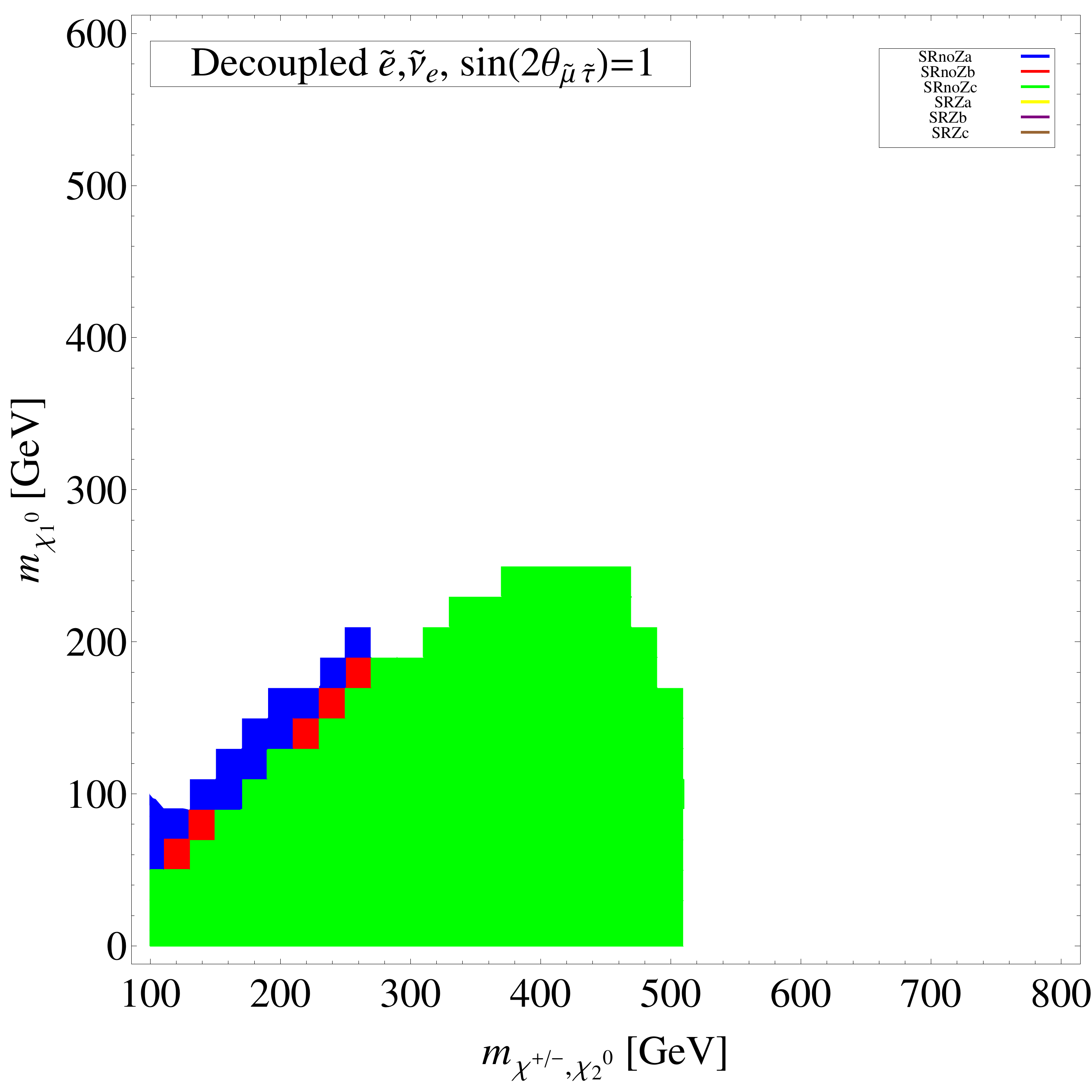}
\label{fig:neut_chi_smuon-stau}
}
\caption{
The excluded region in the chargino-LSP mass plane for 
${\widetilde \ell}_L {\widetilde B}{\widetilde W}$ models
obtained by reproducing the ATLAS search for 
$\ell^+\ell^-\ell^\pm$
with one OSSF pair and missing energy
from chargino-neutralino pair production,
with different assumptions about flavor:
(a) flavor blind sleptons halfway between the chargino and 
LSP masses; (b) same as in (a) but with the selectron and smuon maximally mixed; 
(c) same as in (a) but with the smuon and smuon sneutrino decoupled at 1~TeV; 
(d) same as in (a) but with the selectron and electron sneutrino decoupled at 1~TeV,
and with maximal smuon-stau mixing.
}
\label{fig:neut_chi}
\end{figure}

%
The most sensitive searches in this class of models are based on chargino-neutralino production,
with ${\widetilde\chi}^+{\widetilde\chi}^0$ (${\widetilde\chi}^-{\widetilde\chi}^0$) resulting in three leptons $\ell^+\ell^-\ell^+$ ($\ell^+\ell^-\ell^-$), 
and missing energy. Here, as usual, $\ell=e,\mu$.
Note that one OS lepton pair originates from the neutralino decay, with the third lepton coming from
the chargino.
These signatures were used in~\cite{ATLAS:2013rla}\footnote{Much stronger bounds were obtained by CMS in \cite{Khachatryan:2014qwa}, as 
can be seen in Fig.~\ref{fig:model_G}. However, such an analysis has been not embedded yet in CheckMATE.}, and interpreted in the context of the
${\widetilde \ell}_L {\widetilde B}{\widetilde W}$ models.
Since the slepton spectrum was assumed to be flavor blind, the neutralino decay leads to OSSF
leptons. We reproduce the results for this scenario in Fig.~\ref{fig:neut_chi_flavorblind}.
%
Note that the lepton flavor information is not fully utilized in this analysis.
Rather, apart from the missing energy, the main requirement is OSSF leptons (electrons {\sl or} 
muons), plus a third electron or muon, with different signal regions corresponding to 
the invariant mass of the OSSF pair. As can be seen in 
Fig.~\ref{fig:neut_chi_flavorblind}, the highest sensitivity is obtained
from the SRnoZa, SRnoZb, SRnoZc channels, in which this invariant mass
is required to be far from the $Z$ mass.
Specifically, $m_{ll}<60$~GeV in SRnoZa,  $60<m_{ll}<81.2$~GeV in SRnoZb, 
and $m_{ll}<81.2$~GeV or $m_{ll}>101.2$~GeV in SRnoZc. 
Similarly, other channels (SRZa, SRZb, SRzc) require this invariant mass to be close to the $Z$ 
mass in order to increase sensitivity to chargino or neutralino decays into $Z$ bosons.

We now turn to consider models with flavor dependent sleptons.
Since the search is essentially a counting experiment, targeting three leptons
with charges summing to one,
we expect smuon-selectron mixing to have little effect on the results,
as long as the sleptons are nearly degenerate.
Each slepton mass eigenstate has a 1/6 branching fraction,
independently of the mixing.
Furthermore, any mixed selectron-smuon states would result in
an opposite-sign lepton pair,
with each lepton being either an electron or a muon.
However, a small fraction of the events would have no OSSF pair, 
and would therefore not contribute to the ATLAS signal regions,
leading to a mild reduction in the sensitivity of the search.
This is clearly seen in Fig.~\ref{fig:neutchi_large_mixing}, 
where we show the limit for nearly degenerate, maximally mixed 
selectron-smuon states.
%
The situation would be different in searches looking
for kinematic features, such as the kinematic edge associated with the dileptons
coming from the neutralino decay. In this case, the $\ell^+\ell^-$ flavor is
correlated, and flavor mixing has an important effect~\cite{Galon:2011wh}.
\begin{figure}[t]
\centering
\subfigure[~$\Delta m_{\tilde e\tilde\mu}=5\times10^{-3}m,~\sin2\theta_{\tilde e\tilde\mu}=1$]{
\includegraphics[width=0.47\textwidth]{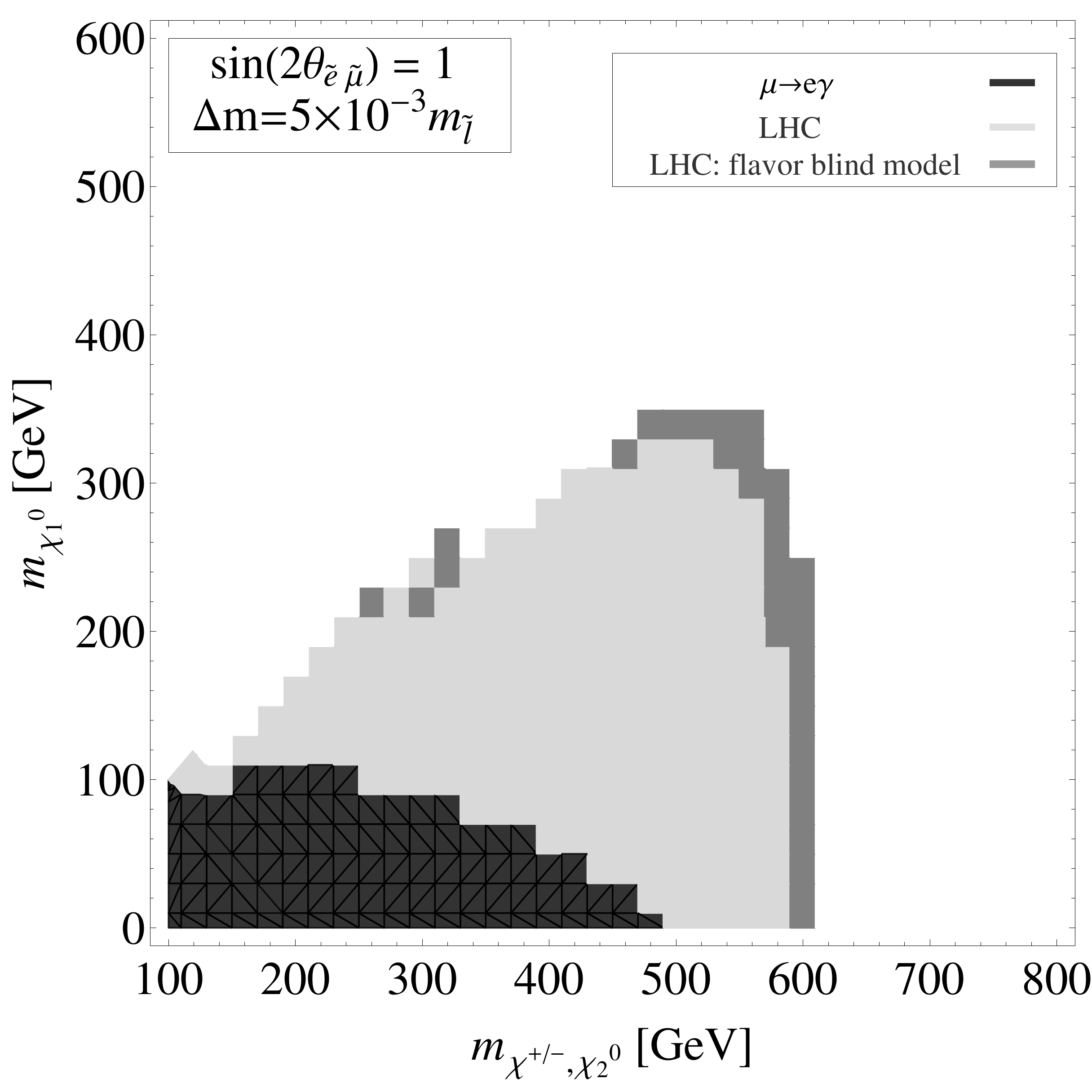}
\label{fig:neut_chi_overlayed_constraints_5sin1_emu}
}
\subfigure[~$\Delta m_{\tilde \mu\tilde\tau}=5\times10^{-3}m,~\sin2\theta_{\tilde \mu\tilde\tau}=1$]{
\includegraphics[width=0.47\textwidth]{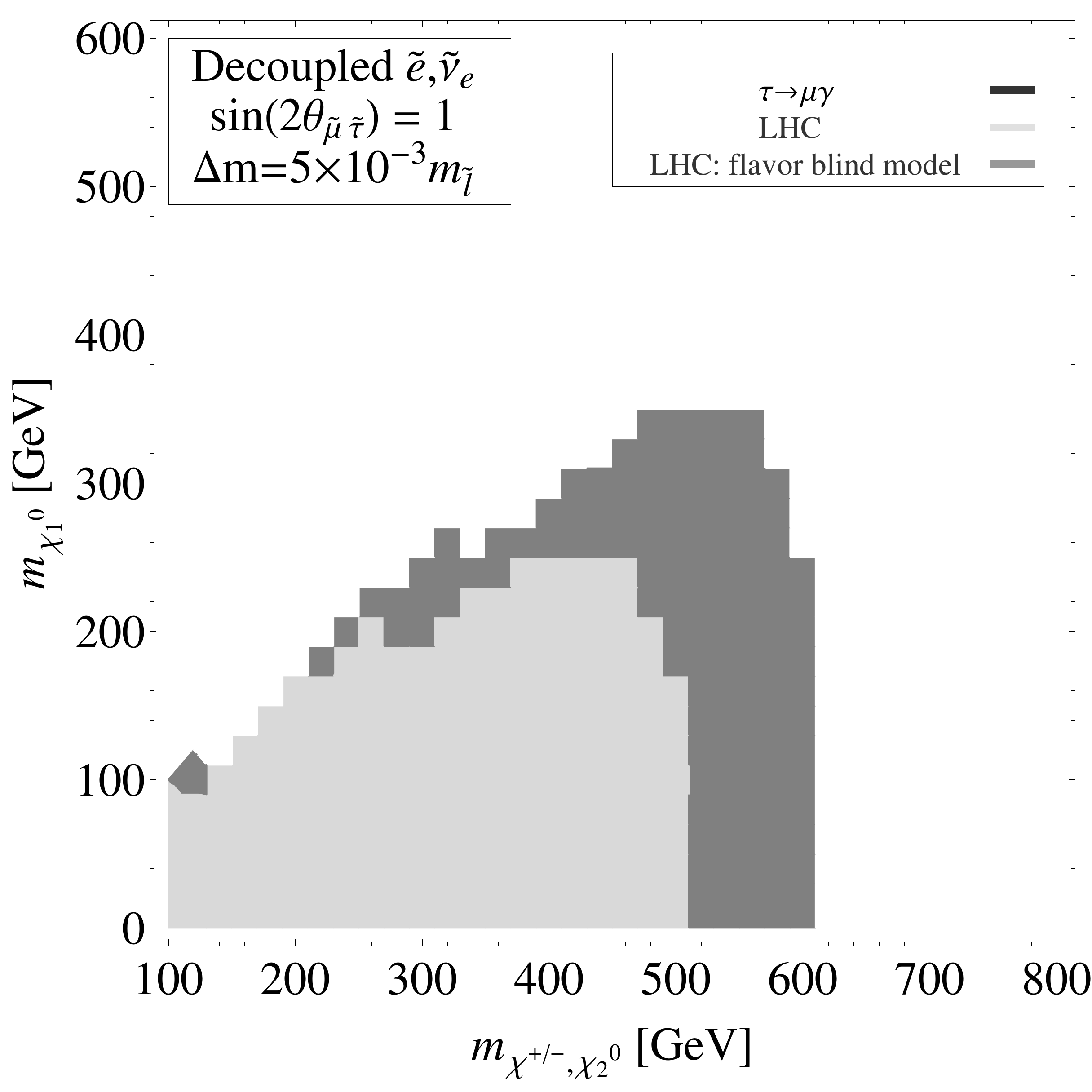}
\label{fig:neut_chi_overlayed_constraints_5sin1_mutau_esplit}
}
\caption{
The excluded region (light grey) 
in the chargino-LSP mass plane for 
${\widetilde \ell}_L {\widetilde B}{\widetilde W}$ models
obtained by reproducing the ATLAS search for 
$\ell^+\ell^-\ell^\pm$
with one OSSF pair and missing energy
from chargino-neutralino pair production,
with different assumptions about flavor:
(a) maximal selectron-smuon mixing with $\Delta m_{\tilde e\tilde\mu}=5\times 10^{-3}m$;
(b) decoupled $\tilde e$, $\tilde \nu_e$, maximal $\tilde\mu-\tilde\tau$ mixing
with $\Delta m_{\tilde\mu\tilde\tau}=5\times 10^{-3} m$.
The dark hatched region is excluded by  $\mu\to e\gamma$.
The dark grey indicates the excluded region of 
Fig.~\ref{fig:neut_chi_flavorblind} (flavor blind sleptons)
which is now allowed.
}
\label{fig:neut_chi_overlayed_constraints}
\end{figure}

A large mass splitting between the selectron and smuon would have a much larger effect
on the analysis. If for example, the smuon and its sneutrino are much heavier than the remaining
sleptons, the branching fraction into taus increases at the expense of electrons and muons.
The reduced reach is clearly seen in Fig.~\ref{fig:neut_chi_split}, where the smuon and muon sneutrino
are decoupled at 1~TeV.
%
Note that, if lepton flavor information were kept, with electrons and muons 
treated separately,
purely electron trileptons would still be sensitive to the presence of the
selectron and the electron sneutrino, which have a larger branching fraction
in this case compared to scenarios with three active flavors.

Finally, we consider smuon-stau mixing. Clearly, smuon-stau mixing, 
or stau-selectron mixing,  reduces the sensitivity of these searches. 
However, as long as the slepton
states are close in mass, the number of $\ell^\pm \ell^\mp \ell^+$ (or $\ell^\pm \ell^\mp \ell^-$) trileptons remains essentially the same.
In Fig.~\ref{fig:neut_chi_smuon-stau}, we modify the model by taking 
the selectron (and its sneutrino)
to be very heavy, with, in addition, maximally-mixed smuon-stau states halfway between the LSP and the heavy gauginos.
Indeed, the reach is significantly lower in this case.
Note that this scenario, under the hypothesis of no mixing involving the 
selectron, is only constrained by $\tau\to\mu\gamma$,
but as can be seen in Fig.~\ref{fig:model_G},
this gives no constraint on the flavor parameters at present. 
For completeness, in Fig.~\ref{fig:neut_chi_overlayed_constraints_5sin1_emu}, 
we show the bounds from $\mu\to e\gamma$ (dark hatched region) together with the LHC results.
The excluded region for the flavor blind model is also shown (dark grey) for reference. 
Similarly, Fig.~\ref{fig:neut_chi_overlayed_constraints_5sin1_mutau_esplit} displays the limits on 
models with decoupled selectrons and with smuon-stau mixing, compared to the flavor-blind models.
All of the parameter space is compatible with $\tau\to\mu \gamma$ in this case.
%

\section{Conclusions}\label{sec:conclusions}
Low energy constraints on slepton flavor are often interpreted
in terms of minimal supersymmetric extensions of the standard model,
with all superpartner masses determined by a few parameters.
If there is anything the first LHC run has taught us however, both 
through direct searches and through the measurement of the Higgs mass,
is that we should be wary of minimal theoretical frameworks when
searching for supersymmetry. With this in  mind, we adopted a model
independent approach to analyze the implications of low energy bounds
on flavor dependent spectra, and the LHC signatures of viable
models. 

Since we are interested in scenarios with both small and large slepton mass
differences,
we computed the low-energy flavor-violating (as well as flavor-conserving) 
dipole amplitudes in the mass-eigenstate basis. 
We obtained compact expressions for the amplitudes, 
classified according to the the ways in which the chirality flip is implemented.

Then, with the aim of comparing the sensitivity of CLFV and LHC 
experiments to flavor dependent slepton spectra, we systematically 
classified the simplified
models involving only non-colored superpartners, 
with at most three different mass scales.
Some of these models were employed by the LHC collaborations to 
interpret their searches for electroweak production of sleptons, 
neutralinos and charginos.
The quantity constrained by CLFV is schematically
given by,
\beq
\frac1{\widetilde{m}^2}\,
\delta\sim
\frac1{\widetilde{m}^2}\,
 \frac{\Delta\widetilde{m}}{\widetilde{m}}\,
\sin2\theta\,.
\eeq
For each of the simplified models, we derived the CLFV bound on $\delta$
for a small mass splitting $\Delta\widetilde{m}$,
 as a function of the superpartner masses,
 showing also the current LHC limits on flavor-blind sleptons,
and, whenever possible,  the projected 14 TeV 
bounds.
As is evident from our plots of section IV, there is an interesting interplay 
between low-energy and high-energy bounds.
In particular, since the cross-sections for superpartner production 
fall very fast with the superpartner mass,
whereas the contributions of heavy superpartners to CLFV processes 
decouple more slowly, low-energy channels
are the best probes of CLFV for heavy spectra. Moreover, LHC bounds are 
rather loose for compressed spectra, in contrast to CLFV processes.

We then turned to LHC searches.
Using our results for the allowed flavor dependence, we considered a few 
simplified
models with either large slepton mass differences or large
selectron-smuon or smuon-stau mixing.
We reinterpreted several ATLAS analyses to obtain  the allowed regions
in the parameter space of these models.
In some cases, flavor dependence significantly modifies the reach of the 
searches.
Since the next LHC run will probe regions in which larger 
flavor effects are allowed, it is important
that full flavor information is retained 
in the experimental analyses,
and that future searches are interpreted taking into account 
the possible large flavor mixing
in the slepton sector.

\section*{Acknowledgments}
IG thanks David Cohen, the ATLAS-Technion computer grid administrator, 
for extensive computer assistance. IG also thanks Jamie Tattersall
for useful discussions.
LC and PP thank the Technion Center for Particle Physics for
hospitality and financial support while some of this work was completed.
The research of YS  was supported by
the Israel Science Foundation (ISF) under grant No.~1367/11, 
by the United States-Israel
Binational Science Foundation (BSF) under grant No.~2010221,
and by the ICORE
Program of the Planning and Budgeting Committee (Grant No.~1937/12).
IG is supported by NSF grant PHY-1316792.
AM gratefully acknowledges support by the research grant Theoretical Astroparticle Physics No. 2012CPPPYP7 under the program PRIN 2012 funded by the Ministero dell’Istruzione, Universita’ e della Ricerca (MIUR). The research of AM and PP is also supported by the ERC Advanced Grant No. 267985 (DaMeSyFla) and by the research grant TAsP (Theoretical Astroparticle Physics), by the Istituto Nazionale di Fisica Nucleare (INFN).

\appendix

\section{Expressions for the leptonic dipoles}
We now present the full expressions for the $\ell_i \to \ell_j \gamma$ amplitudes, $(g-2)_\mu$ and $d_e$ distinguishing among the ways in which 
the chirality flip is implemented. In order to simplify the expressions as much as possible, while retaining the salient flavor 
features, we treat $SU(2)$ breaking effects in the chargino/neutralino mass-matrices as perturbations~\cite{Paradisi:2005fk}. 
On the other hand, we work in the mass eigenstate basis for the sleptons assuming a two generation scheme.

\subsection{$\ell_i \to \ell_j \gamma$}

In the following, we give the relevant amplitudes for 
$\ell_i\to \ell_j \gamma$ both in the mass eigenstate 
basis and in the MIA.
As a first class of contributions, we consider the 
amplitudes in which the chirality flip is realized 
on the external fermion line. In this case we have,
\bea
(A^{c_1}_{L})_{\mysmall SU(2)} & = &
-\frac{\alpha_{2}}{8\pi}
\sin\theta_L\cos\theta_L
\left[
\frac{f^{L}_{c}(x_{\mysmall 2\widetilde\ell_{1}})}{m^2_{\widetilde\ell_{1}}} -
\frac{f^{L}_{c}(x_{\mysmall 2\widetilde\ell_{2}})}{m^2_{\widetilde\ell_{2}}}
\right]\,,
\\
(A^{n_1}_{L})_{\mysmall SU(2)} & = &
\frac{\alpha_{2}}{16\pi}
\sin\theta_L\cos\theta_L
\left[
\frac{f^{L}_{n}(x_{\mysmall 2\widetilde\ell_{1}})}{m^2_{\widetilde\ell_{1}}} -
\frac{f^{L}_{n}(x_{\mysmall 2\widetilde\ell_{2}})}{m^2_{\widetilde\ell_{2}}}
\right]\,,
\\
(A^{n_1}_{L})_{\mysmall U(1)} & = &
\frac{\alpha_{Y}}{16\pi}
\sin\theta_L\cos\theta_L
\left[
\frac{f^{L}_{n}(x_{\mysmall 1\widetilde\ell_{1}})}{m^2_{\widetilde\ell_{1}}} -
\frac{f^{L}_{n}(x_{\mysmall 1\widetilde\ell_{2}})}{m^2_{\widetilde\ell_{2}}}
\right]\,,
\\
(A^{n_1}_{R})_{\mysmall U(1)} & = &
\frac{\alpha_{Y}}{4\pi}
\sin\theta_R\cos\theta_R
\left[
\frac{f^{L}_{n}(x_{\mysmall 1\widetilde e_{1}})}{m^2_{\widetilde e_{1}}} -
\frac{f^{L}_{n}(x_{\mysmall 1\widetilde e_{2}})}{m^2_{\widetilde e_{2}}}
\right]\,,
\eea
where we have defined the ratios $x_{I\widetilde\ell_j} = |M^2_I|/m^2_{\widetilde \ell_j}$, 
where $M_1$ ($M_2$) is the Bino (Wino) mass parameter, $x_{I\widetilde e_j} = |M^2_I|/m^2_{\widetilde e_j}$, 
$x_{\mu\widetilde\ell_j}= |\mu|^2/m^2_{\widetilde \ell_j}$, and $x_{\mu\widetilde e_j}= |\mu|^2/m^2_{\widetilde e_j}$.

The corresponding MIA expressions read
\begin{align}
(A^{c_1}_L)^{\mysmall\rm MIA}_{\mysmall SU(2)} & = \frac{\alpha_2}{4 \pi} \frac{\Delta_{LL}^{21}}{m_L^4}
f_{1c}(x_{\mysmall 2L})\,,
\\
(A^{n_1}_L)^{\mysmall\rm MIA}_{\mysmall SU(2)} & =
\frac{\alpha_2}{4 \pi}\frac{\Delta_{LL}^{21}}{m_L^4} f_{1n} (x_{\mysmall2L})\,,
\\
(A^{n_1}_L)^{\mysmall\rm MIA}_{\mysmall U(1)} & = \frac{\alpha_Y}{4 \pi} \frac{\Delta_{LL}^{21}}{m_L^4}
f_{1n} (x_{\mysmall 1L})\,,
\\
(A^{n_1}_R)^{\mysmall\rm MIA}_{\mysmall U(1)} & = \frac{\alpha_Y}{\pi} \frac{\Delta_{RR}^{21}}{m_R^4}
f_{1n} (x_{\mysmall 1R})\,.
\end{align}
If the chirality flip occurs on the Yukawa vertex, we have the following amplitudes,
\bea
(A^{c_2}_{L})_{\mysmall SU(2)} &=&
\frac{\alpha_{2}}{8\pi}
\sin\theta_L\cos\theta_L
\bigg[
a_2\,\frac{f^{LR}_{c}(x_{2\widetilde \ell_{1}})}{m^2_{\widetilde \ell_{1}}} -
b_2\,\frac{f^{LR}_{c}(x_{\mu\widetilde\ell_{1}})}{m^2_{\widetilde \ell_{1}}}
-({\widetilde \ell_{1}} \leftrightarrow {\widetilde \ell_{2}})
\bigg]\,,
\\
(A^{n_2}_{L})_{\mysmall SU(2)} &=&
-\frac{\alpha_{2}}{8\pi}
\sin\theta_L\cos\theta_L
\bigg[
a_2\,\frac{f_{3n}(x_{2\widetilde \ell_{1}})}{m^2_{\widetilde \ell_{1}}}
-b_2\,\frac{f_{3n}(x_{\mu\widetilde\ell_{1}})}{m^2_{\widetilde \ell_{1}}}
-({\widetilde \ell_{1}} \leftrightarrow {\widetilde \ell_{2}})
\bigg]\,,
\\
(A^{n_2}_{L})_{\mysmall U(1)} &=&
\frac{\alpha_{Y}}{8\pi}
\sin\theta_L\cos\theta_L
\bigg[
a_1\,\frac{f_{3n}(x_{1\widetilde \ell_{1}})}{m^2_{\widetilde \ell_{1}}}
-b_1\,\frac{f_{3n}(x_{\mu\widetilde\ell_{1}})}{m^2_{\widetilde \ell_{1}}}
-({\widetilde \ell_{1}} \leftrightarrow {\widetilde \ell_{2}})
\bigg]\,,
\\
(A^{n_2}_{R})_{\mysmall U(1)} &=&-\frac{\alpha_{Y}}{4\pi}
\sin\theta_R\cos\theta_R
\bigg[
a_1\,\frac{f_{3n}(x_{1\widetilde e_{1}})}{m^2_{\widetilde e_{1}}}
-b_1\,\frac{f_{3n}(x_{\mu\widetilde e_{1}})}{m^2_{\widetilde e_{1}}}
-({\widetilde e_{1}} \leftrightarrow {\widetilde e_{2}})
\bigg]\,,
\eea
where we have defined the quantities
\begin{align}
a_1 &= \frac{(|M_1|^2 + \mu M_1 t_\beta)}{|M_1|^2 - |\mu|^2}\,,\qquad
a_2 = \frac{(|M_2|^2 + \mu M_2 t_\beta)}{|M_2|^2 - |\mu|^2}\,,
\\
b_1 &= \frac{(|\mu|^2 + \mu M_1 t_\beta)}{|M_1|^2 - |\mu|^2}\,,\qquad~~\,
b_2 = \frac{(|\mu|^2 + \mu M_2 t_\beta)}{|M_2|^2 - |\mu|^2}\,.
\end{align}
The corresponding MIA amplitudes are
\begin{align}
(A^{c_2}_L)^{\mysmall\rm MIA}_{\mysmall SU(2)} & =  \frac{\alpha_2}{4 \pi} \frac{\Delta^{21}_{LL}}{m_L^4}
\left[ a_2 f_{2c}(x_{2L}) - b_2 f_{2c} (x_{\mu L}) \right]\,,
\\
(A^{n_2}_L)^{\mysmall\rm MIA}_{\mysmall SU(2)} & =
\frac{\alpha_2}{4 \pi}\frac{\Delta^{21}_{LL}}{m_L^4}
\left[a_2 f_{2n} (x_{2L}) - b_2 f_{2n} (x_{\mu L}) \right]\,,
\\
(A^{n_2}_L)^{\mysmall\rm MIA}_{\mysmall U(1)} & =
- \frac{\alpha_Y}{4 \pi} \frac{\Delta^{21}_{LL}}{m_L^4}
\left[a_1 f_{2n} (x_{1L}) - b_1 f_{2n} (x_{\mu L}) \right]\,,
\\
(A^{n_2}_R)^{\mysmall\rm MIA}_{\mysmall U(1)} & =
\frac{\alpha_Y}{2 \pi} \frac{\Delta^{21}_{RR}}{m_R^4}
\left[a_1 f_{2n} (x_{1R}) - b_1 f_{2n}(x_{\mu R}) \right]\,.
\end{align}
Finally, the amplitudes corresponding to a chirality flip on the internal sfermion line read
\bea
(A^{n_3}_L)_{\mysmall U(1)} & = & -\frac{\alpha_{Y}}{4\pi}
\frac{M_1}{m_{\mu}}
\Delta^{22}_{RL}\sin\theta_L\cos\theta_L
\left[\left(\frac{f_{3n}(x_{1R})}{m^2_{R_2}} - \frac{f_{3n}(x_{1\widetilde\ell_1})}{m^2_{\widetilde\ell_1}}\right)
\frac{1}{( m^2_{R_2} - m^2_{\widetilde\ell_{1}} )} 
- (\widetilde\ell_1 \leftrightarrow \widetilde\ell_2)\right]\nonumber\\
&-& \frac{\alpha_Y}{4 \pi} \frac{M_1}{m_{\mu}} \frac{\Delta^{21}_{RL}}{m_{\widetilde\ell_{1}}^2 - m_{R_2}^2}
\left[\frac{f_{3n}(x_{1{\widetilde\ell_{1}}})}{m_{\widetilde\ell_{1}}^2}-\frac{f_{3n}(x_{1R})}{m_{R_2}^2}\right]\,,
\\
(A^{n_3}_R)_{\mysmall U(1)} & = & -\frac{\alpha_{Y}}{4\pi}
\frac{M_1}{m_{\mu}}
\Delta^{22}_{LR}\sin\theta_R\cos\theta_R
\left[\left(\frac{f_{3n}(x_{1L})}{m^2_{L_2}} - \frac{f_{3n}(x_{1\widetilde e_1})}{m^2_{\widetilde e_1}}\right)
\frac{1}{( m^2_{L_2} - m^2_{\widetilde e_{1}} )} - (\widetilde e_1 \leftrightarrow \widetilde e_2)\right]\nonumber\\
&-& \frac{\alpha_Y}{4 \pi} \frac{M_1}{m_{\mu}} \frac{\Delta^{21}_{LR}}{m_{\widetilde e_{1}}^2 - m_{L_2}^2}
\left[\frac{f_{3n}(x_{1{\widetilde e_{1}}})}{m_{\widetilde e_{1}}^2} - \frac{f_{3n}(x_{1L})}{m_{L_2}^2} \right]\,,
\eea
where $\Delta^{22}_{RL} = m_{\mu}(A_\mu - \mu^* t_\beta)$.

The MIA expressions for this last case are the following
\begin{align}
(A^{n_3}_L)^{\mysmall\rm MIA}_{\mysmall U(1)} & = \frac{\alpha_Y}{4 \pi} \frac{M_1}{m_{\mu}}
\frac{\Delta^{22}_{RL}\Delta^{21}_{LL}}{(m_L^2 - m_R^2)}
\left[ \frac{2 f_{2n}(x_{1L})}{m_L^4} + \frac{1}{m_L^2 - m_R^2}
\left( \frac{f_{3n} (x_{1L})}{m_L^2} - \frac{f_{3n} (x_{1R})}{m_R^2} \right) \right]
\nonumber \\
& -  \frac{\alpha_Y}{4 \pi} \frac{M_1}{m_{\mu}}  \frac{\Delta^{21}_{RL}}{m_L^2 - m_R^2}
\left[ \frac{f_{3n} (x_{1L})}{m_L^2} - \frac{f_{3n} (x_{1R})}{m_R^2} \right]\,,
\end{align}
\begin{align}
(A^{n_3}_R)^{\mysmall\rm MIA}_{\mysmall U(1)} & = \frac{\alpha_Y}{4 \pi} \frac{M_1}{m_{\mu}}
\frac{\Delta^{22}_{LR}\Delta^{21}_{RR}}{(m_R^2 - m_L^2)}
\left[ \frac{2 f_{2n}(x_{1R})}{m_R^4} + \frac{1}{m_R^2 - m_L^2}
\left( \frac{f_{3n} (x_{1R})}{m_R^2} - \frac{f_{3n} (x_{1L})}{m_L^2}\right)
\right]
\nonumber \\
& -  \frac{\alpha_Y}{4 \pi} \frac{M_1}{m_{\mu}}  \frac{\Delta^{21}_{LR}}{m_R^2 - m_L^2}
\left[ \frac{f_{3n} (x_{1R})}{m_R^2} - \frac{f_{3n} (x_{1L})}{m_L^2} \right]\,.
\end{align}
\subsection{$(g-2)_\mu$}
The supersymmetric effects for $\Delta a_{\mu} =(g-2)_\mu / 2$ are such that $\Delta a_{\mu} = \Delta a^{(n)}_{\mu} + \Delta a^{(c)}_{\mu}$ 
where $\Delta a^{(n)}_{\mu}$ and $\Delta a^{(c)}_{\mu}$ arise from the neutralino and chargino contributions, respectively. 
The contributions where the chirality flip is realized on the external fermion line read
\begin{align}
\left(\Delta a^{n_1}_{\mu}\right)^{\!\mysmall R}_{\!U(1)} & =  -   \frac{\alpha_Y}{ 2 \pi}  \frac{m_{\mu}^2}{m_R^2} \,f_n^L(x_{1R})\,,
\\
\left(\Delta a^{n_1}_{\mu}\right)^{\!\mysmall L}_{\!U(1)} & =  -   \frac{\alpha_Y}{ 8 \pi}  \frac{m_{\mu}^2}{m_L^2} \,f_n^L(x_{1L})\,,
\\
\left(\Delta a^{n_1}_{\mu}\right)_{\!SU(2)} & =  - \frac{\alpha_2 }{ 8 \pi} \frac{m_{\mu}^2}{m_L^2} f_n^L (x_{2L})\,,
\\
\left(\Delta a^{c_1}_{\mu}\right)_{\!SU(2)} & = \frac{\alpha_2}{ 4 \pi} \frac{m_{\mu}^2}{m_L^2} f^L_c (x_{2L})\,,
\end{align}
while, in the case where the chirality flip is realized at the Yukawa vertex, we find
\begin{align}
\left(\Delta a^{n_2}_{\mu}\right)^{\!\mysmall R}_{\!U(1)} & =   \frac{\alpha_Y}{ 2 \pi}  \frac{m_{\mu}^2}{m_R^2} 
\left[ {\rm Re}(a_1) f_{3n} (x_{1R}) - {\rm Re}(b_1) f_{3n} (x_{\mu R}) \right]\,,
\\
\left(\Delta a^{n_2}_{\mu}\right)^{\!\mysmall L}_{\!U(1)} & = -\frac{\alpha_Y}{ 4 \pi} \frac{m_{\mu}^2}{m_L^2} \left[ {\rm Re}(a_1) f_{3n} (x_{1L}) - {\rm Re}(b_1)  f_{3n} (x_{\mu L})  \right]\,,   
\\
\left(\Delta a^{n_2}_{\mu}\right)_{\!SU(2)} & = \frac{\alpha_2 }{ 4 \pi} \frac{m_{\mu}^2}{m_L^2} \left[ {\rm Re}(a_2) f_{3n} (x_{2L}) -  {\rm Re}(b_2) f_{3n} (x_{\mu L}) \right]\,,   
\\
\left(\Delta a^{c_2}_{\mu}\right)_{\!SU(2)}  & = -\frac{\alpha_2}{ 4 \pi} \frac{m_{\mu}^2}{m_L^2} 
\left[ {\rm Re}(a_2) f_c^{LR} (x_{2L}) - {\rm Re}(b_2) f_c^{LR} (x_{\mu L}) \right]\,.
\end{align}
Finally, the amplitude relative to a chirality flip at the internal sfermion line is given by
\begin{align}
\left(\Delta a^{n_3}_{\mu}\right)_{\!U(1)}  & = 
\frac{\alpha_Y}{ 2 \pi} \frac{m_{\mu}}{m_L^2 - m_R^2} {\rm Re}(M_1 m_{LR}^2)_{22} \left[ \frac{f_{3n}(x_{1L})}{m_L^2} -
\frac{f_{3n}(x_{1R})}{m_R^2} \right]\,.
\end{align}
\subsection{Electron EDM}
\label{eq:edm}
The supersymmetric effects for the electron EDM $d_e$ are given by $d_e = d_e^{(n)} + d_e^{(c)}$ where
$d_e^{(n)}$ and $d_e^{(c)}$ arise from the neutralino and chargino contributions, respectively.
In contrast to the $g-2$ and $\mu\to e\gamma$ contributions, $d_e$ does not receive contributions 
from a chirality  flip implemented on the external fermion line, as the resulting amplitude is real.
The amplitudes arising from a chirality flip at the Yukawa vertex read
\begin{align}
\left( \frac{d_e^{n_2}}{e}\right)^{\!\mysmall R}_{\!\!U(1)} & = \frac{\alpha_Y}{ 4 \pi} \frac{m_{e}}{m_R^2} \frac{{\rm Im}(\mu M_1)}{M_1^2 - \mu^2} t_\beta 
\left[ f_{3n}(x_{1R}) - f_{3n} (x_{\mu R}) \right]\,,
\\
\left( \frac{d_e^{n_2}}{e}\right)^{\!\mysmall L}_{\!\!U(1)} & = - \frac{\alpha_Y}{ 8 \pi} \frac{m_{e}}{m_L^2} \frac{{\rm Im} (\mu M_1)}{M_1^2 - \mu^2} t_\beta
\left[ f_{3n}(x_{1L}) - f_{3n}(x_{\mu L}) \right] \,,
\\
\left( \frac{d_e^{n_2}}{e}\right)_{\!\!SU(2)} & = \frac{\alpha_2 }{ 8 \pi} \frac{m_{e}}{m_L^2} \frac{ {\rm Im}(\mu M_2)}{M_2^2 - \mu^2} t_\beta
\left[ f_{3n} (x_{2L}) -  f_{3n}(x_{\mu L}) \right]\,,
\\
\left( \frac{d_e^{c_2}}{e}\right)_{\!\!SU(2)} & =   
- \frac{\alpha_2}{8\pi} \frac{m_{e}}{m_L^2} \frac{{\rm Im} (\mu M_2)}{M_2^2-\mu^2} t_\beta \left[ f_c^{LR} (x_{2L}) -  f_c^{LR} (x_{\mu L})  \right]\,,
\end{align}
while those from a chirality flip at the internal sfermion line are given by
\begin{align}
\left( \frac{d_e^{n_3}}{e}\right)_{\!\!U(1)} & =  \frac{\alpha_Y}{ 4 \pi} \frac{{\rm Im} (M_1 m_{LR}^2)_{11}}{m_L^2 - m_R^2}
\left[ \frac{f_{3n}(x_{1L})}{m_L^2} - \frac{f_{3n}(x_{1R})}{m_R^2}  \right] \,.
\end{align}

\subsection{Loop functions}

In this appendix we report the explicit expressions for the loop functions:
\bea
f_{1n}(x)&=&\frac{-17x^3+9x^2+9x-1+6x^2(x+3)\ln x}{24(1-x)^5}\,,\\
f_{2n}(x)&=&\frac{-5x^2+4x+1+2x(x+2)\ln x}{4(1-x)^4}\,,\\
f_{3n}(x)&=&\frac{1+2x\ln x-x^2}{2(1-x)^3}\,,\\
f_{1c}(x)&=&\frac{-x^3-9x^2+9x+1+6x(x+1)\ln x}{6(1-x)^5}\,,\\
f_{2c}(x)&=&\frac{-x^2-4x+5+2(2x+1)\ln x}{2(1-x)^4}\,,\\
f^L_n (x)&=&\frac{1-6x+3x^2+2x^3-6x^2 \log x}{6 (1-x)^4}\,, \\
f^L_c (x)&=&\frac{2+3x-6x^2+x^3 + 6 x \log x}{6 (1-x)^4}\,, \\
f^{LR}_c (x)&=&\frac{-3 + 4 x - x^2- 2  \log x}{ (1-x)^3}\,.
\eea

\nocite{*}
\providecommand{\href}[2]{#2}\begingroup\raggedright\endgroup

\end{document}